\documentclass[preprint,12pt,titlepage,aps,showpacs]{revtex4}
\def\beq{\begin{equation}}
\def\eeq{\end{equation}}
\def\bea{\begin{eqnarray}}
\def\eea{\end{eqnarray}}

\usepackage{graphicx}

\begin{document}

\title{Quantum Monte Carlo Studies of Superfluid Fermi Gases}
\stepcounter{mpfootnote}
\author {S. Y. Chang}
\affiliation { Department of Physics,  
  University of Illinois at Urbana-Champaign,
        1110 W. Green St., Urbana, IL 61801, U.S.A.}
\author {J. Carlson}
\affiliation { Theoretical Division, 
     Los Alamos National Laboratory,  
      Los Alamos, NM 87545, U.S.A. }	
\author {V. R. Pandharipande}
\affiliation { Department of Physics,  
  University of Illinois at Urbana-Champaign,
        1110 W. Green St., Urbana, IL 61801, U.S.A.}
\author {K. E. Schmidt}
\affiliation {Department of Physics and Astronomy, Arizona State University, Tempe,
AZ 85287, U.S.A.}

\date{\today}

\begin{abstract}

We report results of quantum Monte Carlo calculations of the ground state of dilute Fermi gases
with attractive short range two-body interactions. The strength of the interaction is varied to 
study different pairing regimes which are characterized by the
product of the $s$-wave scattering length and the Fermi wave vector,
$ak_F$.
We report results for the ground state energy, the pairing gap 
$\Delta$ and the quasiparticle spectrum. In the weak coupling regime, $1/ak_F < -1$, we obtain
BCS superfluid and the energy gap $\Delta$ is much smaller than the Fermi gas energy $E_{FG}$.
 When $ a > 0$, the interaction is strong enough to form bound molecules with energy $E_{mol}$.
For $1/ak_F \gtrsim 0.5$ we find that weakly interacting composite bosons are formed in the superfluid
 gas with $\Delta$ and gas energy per particle approaching $|E_{mol}|/2$. In this region we seem to have Bose-Einstein condensation (BEC) of molecules.
The behavior of the energy and the gap in the BCS to BEC transition region, $-0.5 < 1/ak_F < 0.5$ is
discussed.
\end{abstract}
\maketitle

\section{INTRODUCTION}\label{sec1}
How pairing evolves from the bare interaction has been a major question in condensed matter physics, and 
the study of pairing in relation to the phenomenon of superfluidity and superconductivity
 can be traced back to Cooper et al. \cite{cooper59}. Pairing lies at the core of several
  quantum many-body problems, and it is also believed to influence the evolution of 
  neutron stars \cite{pethick1995}.
Here we report results of quantum Monte Carlo calculations of a superfluid Fermi gas with
short range two-body interactions. The strength of the interaction is varied to 
study different regimes of pairing.

The evolution of pairing with the strength of the interaction has been discussed in the literature
 \cite{leggett1980}, \cite{randeria95}.
In the regime where the interaction is weak and attractive,
a gas of fermions has a superconducting instability at low temperatures,
and a gas of Cooper pairs is formed. The typical coherence length 
is larger than the interparticle spacing $r_0$ ($4\pi r_0^3\rho = 3$ with
$\rho$ the number density) and the bound pairs overlap.
In contrast, in 
the strong-coupling limit the coherence length is small, and the bound pairs can be treated as 
well seperated 
Bose molecules. One then expects the molecules to undergo Bose-Einstein condensation (BEC) into
a single quantum state with zero momentum. 

The Bardeen-Cooper-Schrieffer(BCS) theory \cite{leggett1980} and Gorkov equations \cite{gorkov61} have been used to estimate gaps in
superfluid gases. However, their predictions differ by more than a factor of two and they may be
qualitatively valid only in the weakly interacting regime.
Here we use first principle quantum Monte Carlo methods to study the entire region ranging from
free fermions to the tightly bound Bose molecules.

 Dilute Fermi gases of $^{40}$K, $^6$Li, $^2$H for example, can now be studied in
the laboratory using magnetic and optical trapping and ingenious cooling methods \cite{demarco1999}, \cite{ohara2002}. 
These are dilute Fermi systems, in contrast to dense atomic liquid $^3$He or a solution of $^3$He in superfluid $^4$He.
Within the last few years temperatures $T \ll T_F$ have been achieved, where $T_F = \frac{\hbar^2 k_F^2}{2m}$ is the Fermi
kinetic energy and $k_F$ is the Fermi wave vector. At such a low temperature, the fermionic nature of the quantum statistics becomes evident in the measurement of
the density profile of the trapped gas. 
 At even lower temperatures the transition to the superfluid
Cooper-paired state is expected. However, the temperature $T_c$ of this transition can be much lower than $T_F$ and
conclusive evidence of superfluidity is still to be seen. In order to have the transition at an
achievable temperature, the experimentalists rely on the Feshbach resonance technique to produce strong interaction between the
fermionic atoms. 

When the range of the interatomic interaction is smaller than all the length scales in the system, 
the details of the interaction are believed to be unnecessary and the scattering length $a$ is sufficient to characterize it.
 Near the resonance, the magnitude of the scattering length $a$ becomes much larger than $r_0$
 and the system enters the strong-coupling regime.  The value $a k_F \sim -7.4$ has been achieved by O'Hara
 et al. \cite{ohara2002} and the limit ($a k_F \rightarrow -\infty$) is now approached in the laboratory 
 \cite{stenger1999}, \cite{roberts2001}. Recently, creation of bosonic molecules from $^{40}$K atoms  
 was reported by Regal et al. \cite{regal2003}, and pairing in the $1/ak_F \sim 0$ regime was observed (Ref. \cite{regal2004}).

A few words are in order regarding the language of $s$-wave scattering.
 For a noninteracting system at zero temperature, the only length scale is $1/k_F$. 
 We can use the dimensionless quantity $a k_F$ to describe a dilute gas
having interparticle spacing $r_0$ much greater than the interaction range.
We often use $1/{a k_F }$ because $a k_F$ changes discontinuously from $-\infty$ to $+\infty$ when a
bound state is formed at $1/ak_F = 0$. For attractive interactions $1/a k_F$ can change from large negative values (weakly interacting limit) to large positive 
values
(strongly interacting limit). As discussed in section \ref{sec2}, the radius of the bound molecule 
provides another length scale in the strongly interacting regime.  
Some physical examples of the limits of $1/a k_F$ are: 1) electrons in superconductors have $1/a k_F$ large and negative; 
 2) neutron matter has $1/a k_F$ small and negative; and 3) cold deuterium atoms have large positive $1/a k_F$. In the
 last case, molecular bound states smaller than the average interparticle distance $r_0$ are possible. 
 On the other hand, superfluid $^3$He is not describable in terms of $a k_F$, because the interaction
 range is greater than $r_0$, and the paired state does not have $s$-wave symmetry. 
 
 In the limit of zero energy for the colliding pair, the two-body scattering cross section $\sigma$
 is given by $4\pi a^2$. When $|a| \ll r_0$, the interatomic collisions in the gas are similar to those
 in vacuum, and the mean free path is approximately given by $\ell =\frac{1}{\sigma \rho}$.
 However, this approximation is meaningful only when $|a| \ll r_0$ and $ \ell > r_0$.
 When $ |a|$ is $\gtrsim  r_0$ the two-body collisions in the gas are strongly influenced by the
 presence of other particles, and their cross section in the gas is much smaller than in vacuum.
 
 For a Fermi gas at low density, an expansion of the energy in terms of $a k_F$ is possible.
 For spin $1/2$ Fermi gases it is known to be \cite{lenz1929}, \cite{huang1957}
 \beq
 \frac{E}{N} =  E_{FG} \left[ 1+ \frac{10}{9\pi}(a k_F) + \frac{4}{21}(11- 2 \ln 2)(a k_F)^2
 + {\cal{O}}(a k_F)^3 + \ldots ~\right],
 \label{low_exp}
 \eeq
where $E_{FG} = \frac{3}{5} \frac{\hbar^2 k_F^2}{2m} = \frac{3}{5}T_F$ is the ground state energy per particle of the noninteracting Fermi gas.
In the $ ak_F \rightarrow -\infty$ limit, theoretical estimates of 0.326 and 0.568 $E_{FG}$ were reported \cite{baker1999}, \cite{heiselberg2001}.
 More recently, the authors \cite{carlson2003} predicted $E_0 = (0.44 \pm 0.01)E_{FG}$ using quantum Monte Carlo methods. In this paper we continue that study of the properties of 
 cold dilute spin $1/2$ fermion gas and extend it to all the regimes of $1/a k_F$ as a first step for understanding the superfluidity and the bosonization
of dilute Fermi gases. 
 
The model considered in this study consists of $A$ fermions contained in a box with periodic
conditions on its boundaries. It is not polarized so that half of the spins point up and the other half down.
Typically $A$ is varied from 10 to 20 to estimate properties of uniform gas in the $A \rightarrow \infty$
thermodynamic limit. In some cases larger values of $A$ are used.
Fermions of the same spin do not feel the effects of interaction because it is of short range
and Pauli exclusion predominates. The fermions of different spins interact via a central potential $v(r)$ with
the following properties: 
1) It is attractive with very short range as we assume the dilute limit,
2) The details of the potential do not matter, in principle we can think of it as an attractive delta function potential,
and 3) The potential can be adjusted such that we can sweep through different regimes of $a k_F$.

 From the considerations mentioned above, a $\cosh$ potential of the form
\beq
v(r) = - v_0 \frac{2 \hbar^2}{m} \frac{\mu^2}{\cosh^2(\mu r)}~,
\eeq
can be used. The strength of potential ($v_0$) is adjusted to obtain the desired value of $a k_F$. We can also take appropriate
values of $\mu$ such that the effective range of the potential $R_{eff}$ is much smaller than the
interparticle distance $r_0$. When $v_0 = 1$ this potential has $a = \pm \infty$ and $R_{eff} = 2/\mu$.
In most calculations we have used $\mu r_0 = 12$. For the $a \rightarrow -\infty$ case we also tested 
the $\mu r_0 \rightarrow \infty$ limit using $\mu r_0 = 24$ \cite{carlson2003}. 

Results of simple lowest order constraint variational (LOCV) calculations are reported in
section \ref{sec2}. The LOCV method was first used to study neutron matter \cite{vijay73}.
Recently, Cowell et al. \cite{cowell02} have used it to study cold Bose gases in the unstable 
$a > r_0$ regime. It provides a surprisingly good estimate of the ground state energy.
Here we use it to study the effect of the difference between the $\cosh$ ($\mu r_0 = 12$) and delta-function
potentials on the energy of dilute gases.  The difference becomes significant when $1/ak_F \rightarrow 
\infty$, and the radius of the molecule approaches $1/\mu$.  LOCV is also used to estimate the
energy of the unstable state of the Fermi gas for $a > 0$.  The stability of dilute gases 
is discussed in the LOCV section \ref{sec2}. 

One of the limitations of LOCV is that it can not be used to calculate the pairing energy gap $\Delta$ 
or the other superfluid properties of Fermi gases.
The quantum Monte Carlo methods used in Ref. \cite{carlson2003} and this work 
to study superfluid gases are described in section \ref{sec3},
 and the results for the energy, pairing gap and the quasiparticle spectrum 
 are presented in section
 \ref{sec4} over the range $a k_F = -1$ to $\mp \infty$ to $+0.5$. 
Conclusions are given in the last section \ref{sec5}.

\section{Lowest Order Constraint Variational CALCULATIONS}\label{sec2}

In the lowest order constraint variational (LOCV) method the ground state of the Hamiltonian
\beq
\mbox{$\mathcal{H}$} = - \frac{\hbar^2}{2m} \sum\limits_{p = 1}^{A}  \nabla_p^{
2}  + \sum\limits_{i,j'} v(r_{ij'})~,
\label{locv_h}
\eeq
where the unprimed index $i$ denotes spin up particle, primed index $j'$ denotes spin down particle,
 and $p$ can be any particle, is approximated by the Jastrow-Slater wave function
\beq
|\Psi_V \rangle  =  \prod\limits_{i,j'} f(r_{ij'}) |\Phi_S \rangle~,  
\label{locv_wf}
\eeq
where $ |\Phi_S \rangle$ is the ground state of noninteracting fermions.
In the present case $|\Phi_S \rangle$ is a product of two Slater determinants, 
the first corresponding to the spin up fermions and the second
corresponding to the spin down fermions. The interaction effects are represented by the Jastrow
function $\prod\limits_{i,j'} f(r_{ij'}) $, where $f(r_{ij'})$ denotes the
pair correlation function. We often use $f_{ij'}$ to denote $f(r_{ij'})$.
 $f_{ij'}=1$ means no correlation between the pair $ij'$ and $f_{ij'} \ne 1$ for correlated
 pairs. In variational calculations, the function $f(r)$ is determined by minimizing the
  expectation value of the Hamiltonian
\begin{widetext}
\beq 
\langle \mbox{$\mathcal{H}$} \rangle 
= \frac{-\frac{\hbar^2}{2 m} \sum\limits_p \langle \Phi_S| \prod\limits_{\alpha,\beta'}f_{\alpha \beta'} ~\nabla_p^{ 2} 
\prod\limits_{\gamma,\mu'} f_{\gamma \mu'} |\Phi_S \rangle + \sum\limits_{i,j'} 
\langle \Phi_S| \prod\limits_{\alpha,\beta'} f_{\alpha \beta'} \ v_{ij'} \prod\limits_{\gamma,\mu'} f_{\gamma \mu'}
|\Phi_S \rangle }
{\langle \Phi_S| \prod\limits_{\alpha,\beta'} f_{\alpha \beta'} ~
  \prod\limits_{\gamma,\mu'} f_{\gamma \mu'} |\Phi_S \rangle}~.
  \label{locv_nrg}
\eeq
\end{widetext}

The assumption behind LOCV is that the energy is most sensitive to the correlations of short (less than $r_0$)
range. We impose a constraint on the range of $f(r)$ to assure that the correlations are mostly among 
the closest pairs, and keep only the pair terms in the cluster expansion of the energy expectation value.
The healing distance $d$ is the range of $f(r)$ defined such that
$f( r > d )  = 1$ and $\frac{df(r)}{dr}|_{ r = d }= 0$. In LOCV, $d$ is chosen such that
on average there is only one other particle within the distance $d$ of any particle.
 Effects of deviations from this average are assumed to cancel.

\begin{figure}
\includegraphics[width=\columnwidth]{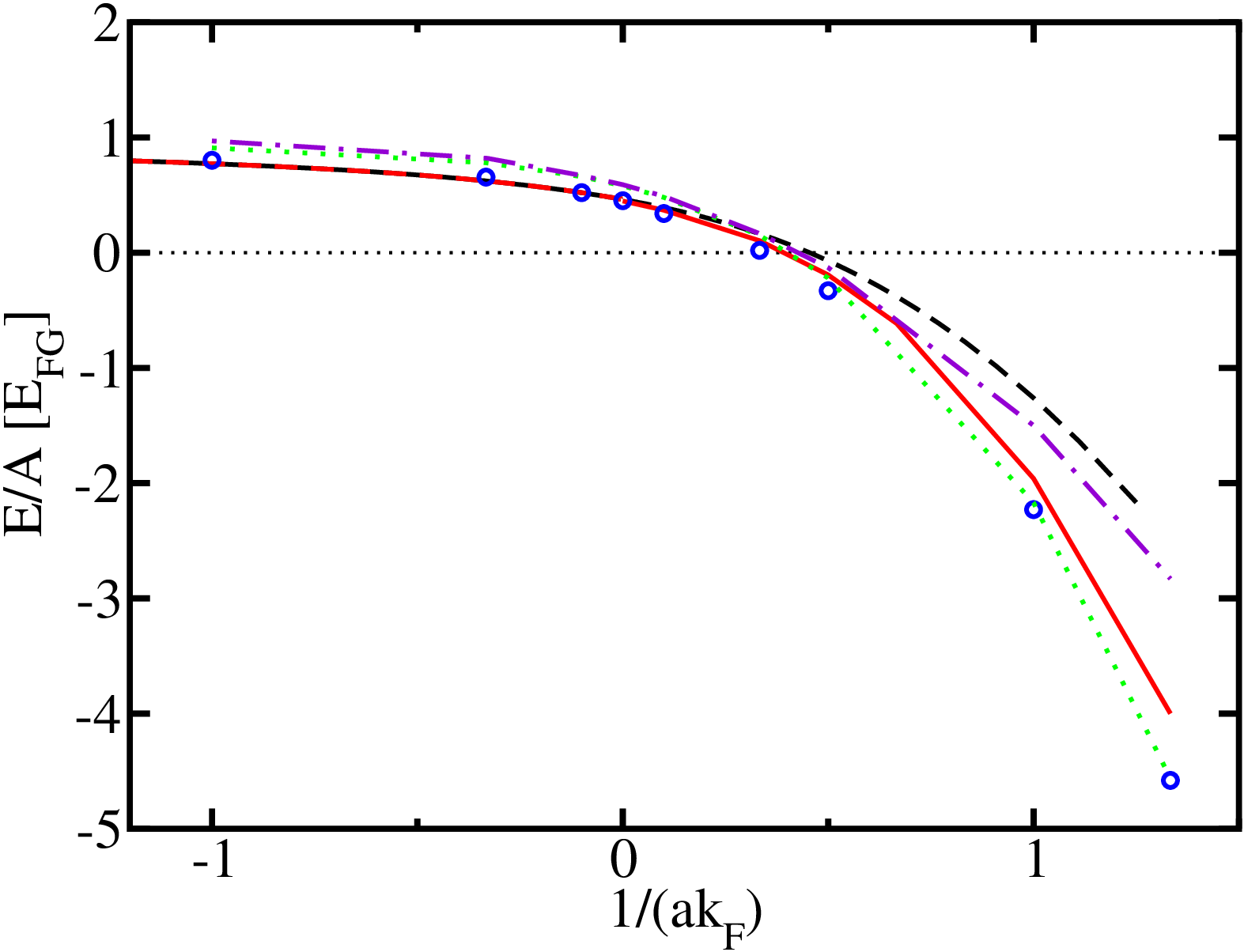}
\caption{Ground state energy per particle of dilute Fermi gases as a function of $a k_F$. The full and
dashed curves give the LOCV results for $\cosh$ ($\mu r_0 = 12$) and delta-function potentials,
and the circles show the essentially exact results for the $\cosh$ potential
obtained with the GFMC method described in section \ref{sec3}. The dotted and dash-dotted curves correspond to 
the conventional BCS results with $\cosh$ and delta-function potentials respectively.}
\label{fig_one}
\end{figure}

Euler-Lagrange minimization of the energy expectation value\cite{kevin77} gives a Schr\"{o}dinger-like equation
for $f(r<d)$
\beq
-\frac{\hbar^2}{m} \nabla^2 f(r) + v(r) f(r) = \lambda f(r)~,
\label{locv_eqn}
\eeq
The constraint used to determine the healing distance is
\beq
\frac{\rho}{2} \int_0^d f^2(r) d^3{\bf{r}} = 1~, 
\label{locv_cond}
\eeq
and the $\lambda$ is chosen such that $\frac{df(r)}{dr}|_{ r = d }= 0$.
In the equations (\ref{locv_eqn}) and (\ref{locv_cond}) we do not have exchange contributions because the range of the interaction is 
short and fermions of same spin do not interact.
When the equations (\ref{locv_eqn}) and (\ref{locv_cond}) are simultaneously solved,
 the energy per particle is given by
\beq
E_{LOCV} = E_{FG} + \frac{\lambda}{2}~.
\eeq
The results obtained for the ground state energy of spin $1/2$ Fermi gas with the $\cosh$ and delta-function
potentials are shown in Fig. \ref{fig_one}.

When $1/ak_F < 0$ the $r_0$ is the only length scale in the
gas, and the results obtained with the $\cosh$ potential with $\mu r_0 = 12$ are indistinguishable from
those given by the delta-function potential. In contrast, when $1/ak_F > 0$, we have a molecular bound state whose
radius provides another length scale. At large positive values of $1/ak_F$ there are differences between 
results of the present $\cosh$ and delta-function potentials due to the {\it rms} radius, $R_{rms}$ of the molecule
becoming comparable to the range of the present $\cosh$ potential. For example, at $1/ak_F = 2$
we get $\mu R_{rms} = 2.3$ with the present choice of $\mu$.
 In principle, we can continue to approximate the delta-function
interaction with the $\cosh$ potential by further increasing $\mu$, and working in the 
$\mu R_{rms} \rightarrow \infty$ limit. However, all of the present 
computations are with $\mu r_0 = 12$.

 Fig.\ref{fig_one} also shows the presumably exact results obtained with the
  $\cosh$ potential with the GFMC method described in the next section. The LOCV
  energies appear to be surprisingly accurate. However, it should be realized
  that a part of the accuracy of LOCV is due to a cancellation of errors, and
  not due to the quality of the Jastrow-Slater variational wave function (Eq.
  \ref{locv_wf}). In fact, the variational energy upper bound obtained with that
  wave function for $1/ak_F = 0$ is $= (0.62 \pm 0.01) E_{FG}$, significantly
  above GFMC result of $ (0.44 \pm 0.01) E_{FG}$. The LOCV energy of $0.46 E_{FG}$
  is below the Jastrow-Slater variational upper bound because it is calculated
  approximately keeping only two-body cluster contributions. However, when the
  contributions of $\geq 3$-body clusters become important we can expect that
  the approximations in the Jastrow-Slater wave function would also become
  important, and the true energy will be below the Jastrow-Slater upper bound.
  
 The ground state energies obtained with the conventional BCS (variational) method are also shown in Fig. \ref{fig_one}.
 In the weakly interacting limit, $1/ak_F \rightarrow -\infty$, the BCS energy is too large since it does not
 have the correct low density limit given by Eq. \ref{low_exp}. On the other hand,
 in the strongly interacting limit, $1/ak_F \rightarrow +\infty$, the BCS energy is very close to the exact result
 (GFMC) presumably because in this limit we have complete pairing of the fermions into Bose molecules.
 LOCV is less accurate than the conventional BCS method in strong coupling region.

\begin{figure}
\includegraphics[width=\columnwidth]{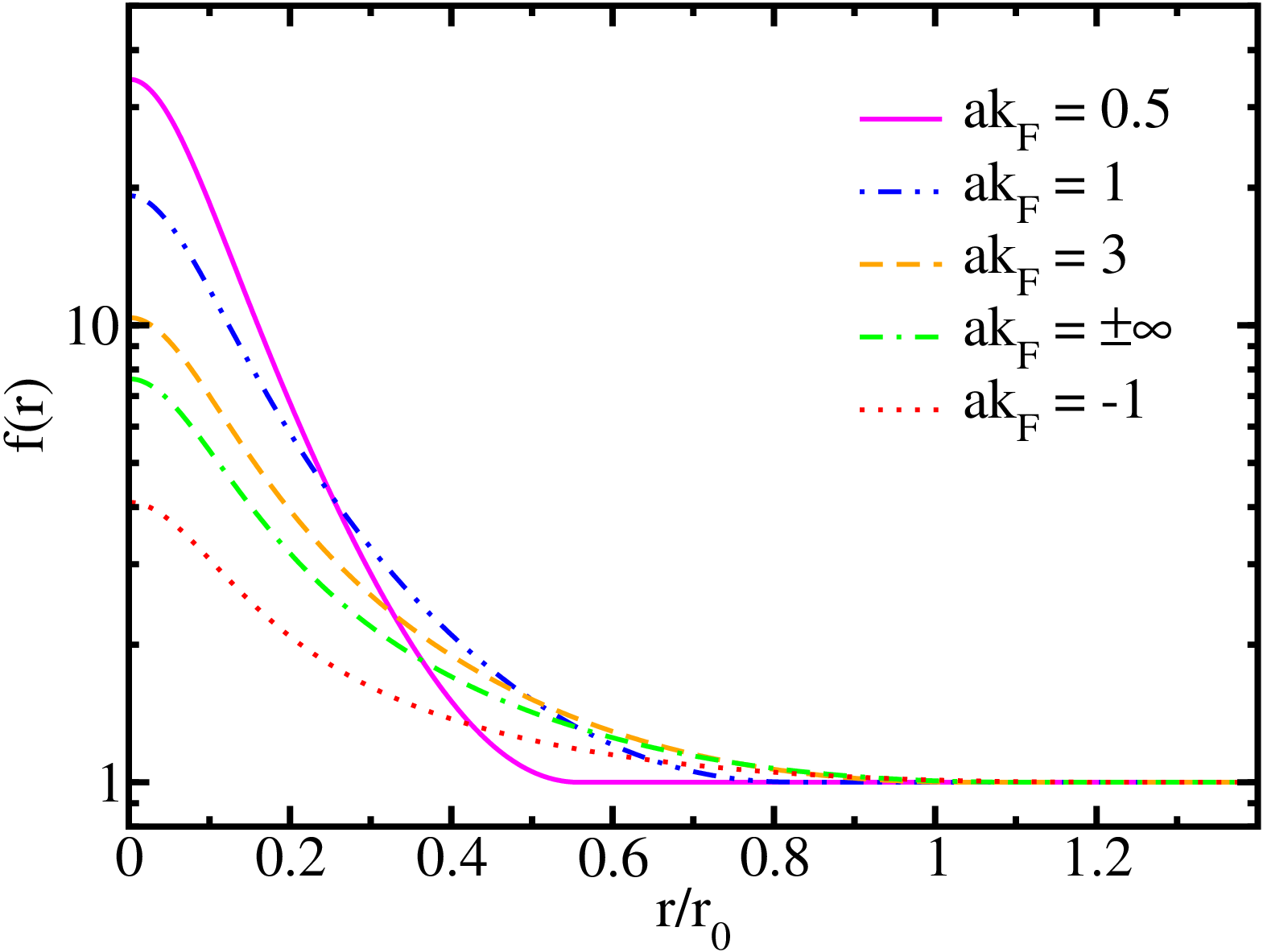}
\caption{Correlation function $f(r)$ for different values of $a k_F$ in the LOCV approximation using
the $\cosh$ potential with $\mu r_0 = 12$.}
\label{fig_two}
\end{figure}

 The LOCV pair correlation functions are shown in Fig. \ref{fig_two}. 
  The healing distance $d \approx r_0$ in the weakly interacting region ($1/ak_F << 0$), and as we increase the strength of the 
 potential, $f(r)$ becomes more and more peaked at the origin, and $d$ becomes smaller than $r_0$.
 In fact for $1/ak_F >> 0$, the boundary condition at $d$ has less impact on $E_{LOCV}$ and
 $\lambda$ of the Eq. \ref{locv_eqn} becomes close to the molecular binding energy $E_{mol}$ such that 
 $E_{LOCV} = E_{FG} + \frac{E_{mol}}{2} + \delta E$. $\frac{E_{mol}}{2}$ is the term that predominates in this limit.
 $\delta E$ is small ($|\delta E| < E_{FG}$) and negative so that $E_{LOCV} > E_{mol}/2$.

\begin{figure}
\includegraphics[width=\columnwidth]{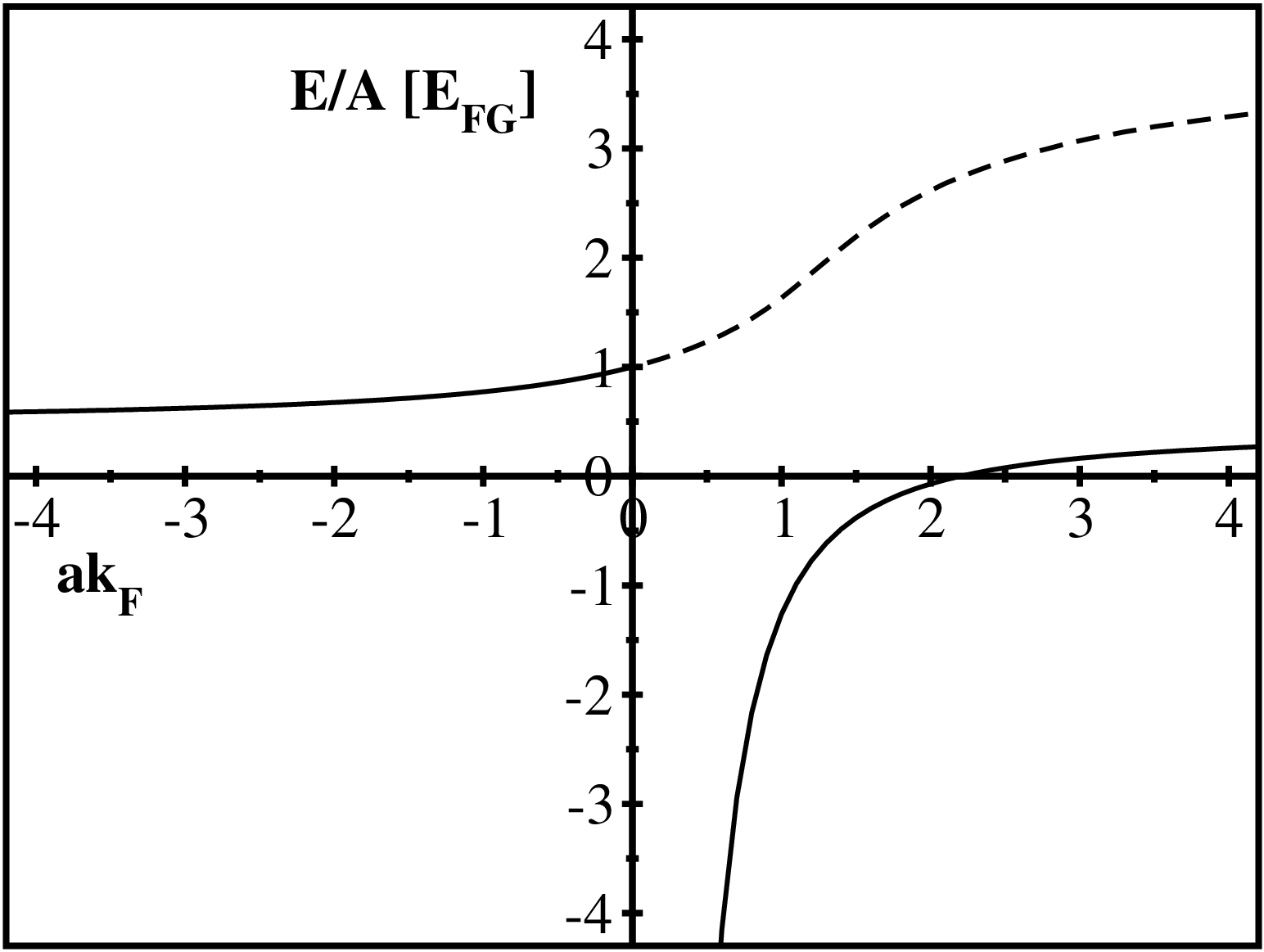}
\caption{The LOCV $E/A$ in units of $E_{FG}$ vs $ak_F$ for attractive delta-function potential. The dashed line
corresponds to $f(r)$ having one node, and the solid line shows the results with nodeless $f(r)$.}
\label{fig_three}
\end{figure}

 When $a > 0$ we can obtain another solution of the LOCV equation with a node at $r < d$. This solution
 was discussed by Cowell et al. \cite{cowell02} for cold Bose gases, and at small values of
 $a k_F$ it gives results in agreement with the low density expansion, (Eq.\ref{low_exp}). The first term
  ($\frac{10}{9\pi}ak_F$) is correctly reproduced by LOCV, but the higher order terms are approximate. In the
  limit $a \rightarrow \infty$ we have the condition $ k d \tan(kd) = -1$ discussed in Ref. \cite{cowell02}.
 The solution with one node is $kd = 2.7983$ and it gives $E/N = E_{FG} + \frac{\lambda}{2} = E_{FG} 
  + \frac{\hbar^2}{2m}\frac{(kd)^2}{d^2} \approx 3.92 E_{FG}$.  Results obtained
  with the delta-function potential, including this unstable region are shown in Fig. \ref{fig_three}.
   Those corresponding to the nodeless solution of the LOCV equation are represented
   by full line, while the dashed line corresponds to the solution with a node.
  
 The state of the gas having a node in the pair correlation function $f(r)$ is 
 unstable because it has 
 energy $>~E_{FG}$, while that with nodeless $f(r)$ has lower energy $<~E_{FG}$ 
 (see Fig. \ref{fig_three}). 
 However, it can have a relatively long life time because energy conservation 
 prevents two atoms to make the transition to the lower energy state.  At least 
 three atoms are needed, which hinders the transition at low densities.  Most of 
 the observed BEC of Bose atoms are in such unstable states in which the 
 $f(r)$ has nodes at small $r$. 
  
  The $E(ak_F)$ shown by the solid line in Fig.\ref{fig_three} corresponds to the stable ground state of the model
  Hamiltonian with the  delta-function interaction. In principle, this state can be exactly calculated
  by the quantum Monte Carlo method described in the next section.
  However, when the range of the interaction is finite, as for the $\cosh$ model,
  the system can collapse to a tightly bound state at large density. This instability can be easily
  seen in the Hartree mean field approximation in which
  \bea
  E_{MF}(\rho) & = & E_{FG}(\rho)  + \frac{\rho}{4} I_v~, \nonumber\\
  I_v & = & \int v(r) d^3{\bf{r}}~,
  \eea
  where $I_v ~(<0)$ is the volume integral of the interaction. At large enough $\rho$ the interaction energy
  becomes larger than $E_{FG}$ leading to a tightly bound state.
  
  Consider for example a simple square well potential of range $R$ such that $v(r<R)  = -V_0$ and $v(r>R) = 0$.
  Let this potential correspond to $a = \infty$. This means $V_0 = \frac{\hbar^2 \pi^2}{4 m R^2}$ and,
  $ I_v     =  -\frac{4\pi}{3} V_0 R^3 =  -\frac{\hbar^2 \pi^3}{3 m} R $. Then
 \beq
 E_{MF}(k_F)  = \frac{3}{5}\frac{\hbar^2 k_F^2}{2m} \left[ 1 - \frac{5 \pi}{54} R k_F \right]~.
 \eeq
  The collapse occurs at values of $k_F > \frac{54}{5\pi} \frac{1}{R}$, and can be pushed to higher densities by reducing
 $R$, or equivalently increasing $\mu$ in the case of the $\cosh$ potential. In the present studies, we ignore this
 collapsed state; assuming that it occurs at too large a density to influence the dilute gas properties.

\section{Green's Function Monte Carlo CALCULATIONS}\label{sec3}
 Green's function Monte Carlo (GFMC) \cite{kalos74} is a powerful method for calculating the
 ground state properties of many-body quantum systems. It can be used to calculate the ground state properties of Bose systems with controllable statistical
 errors without approximation. For the fermion systems, however, we have to deal with the 
 sign problem posed by the anti-symmetry of the wave function as discussed below. We begin with a brief
 overview of the GFMC method.

\begin{figure}
\includegraphics[width=\columnwidth]{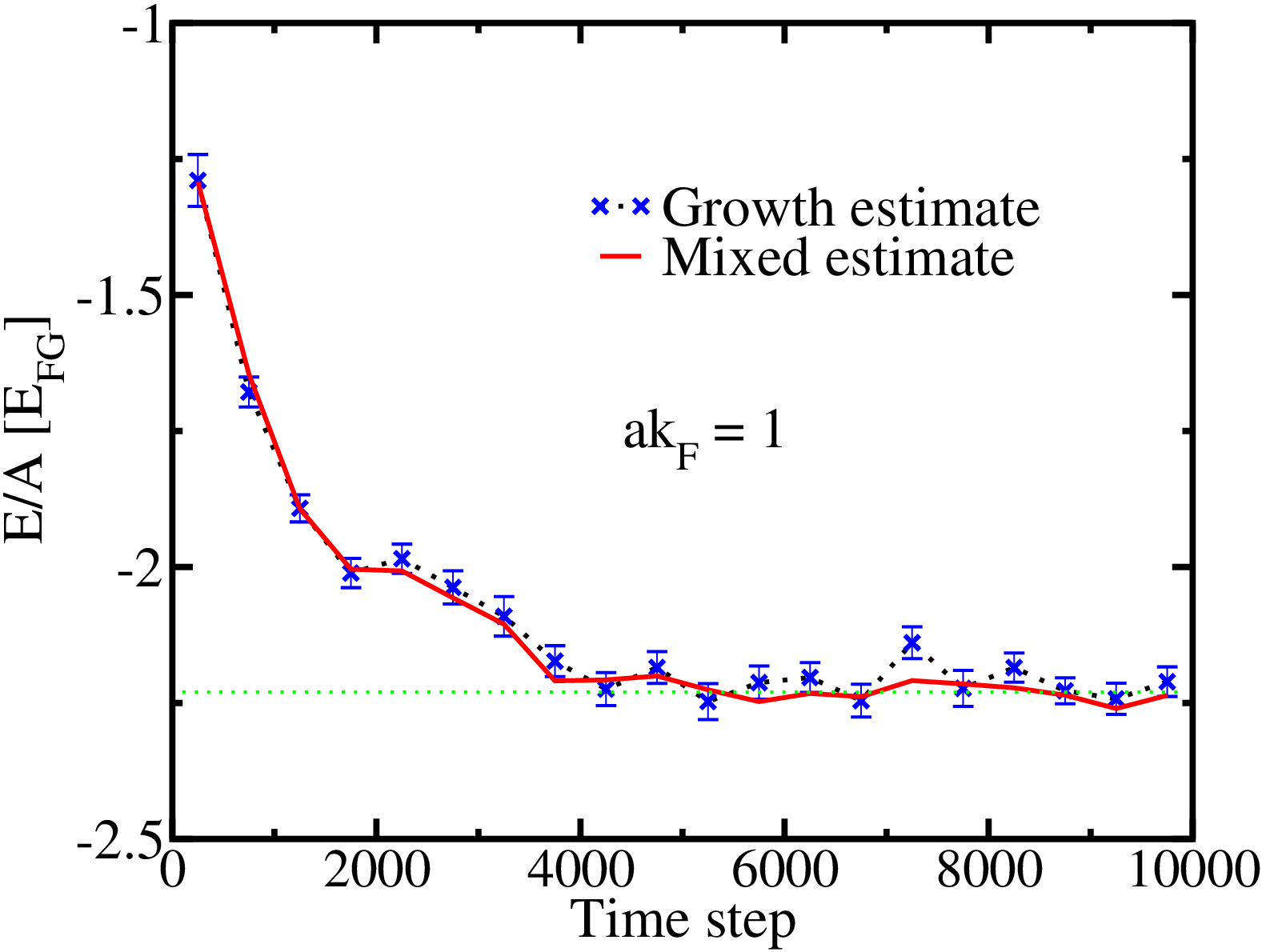}
\caption{Mixed and growth estimates of the GFMC energy. The $\tau \rightarrow \infty$ asymptotic value
is reached after $\sim 5000$ times steps. Each time step $\Delta \tau$ is $1.2 ~ 10^{-3} \frac{\hbar}{E_{FG}}$.}
\label{fig_four}
\end{figure}

Let $\Psi_i$ be the eigenstates of $\cal{H}$ with eigenvalues $E_i$. The trial variational wave function
$\Psi_V$, which provides an approximation to the ground state $\Psi_0$, can be expanded as
\beq
\Psi_V = \sum_i \alpha_i \Psi_i ~.
\eeq
In GFMC we project out $\Psi_0$ from $\Psi_V$ by evolution in imaginary time
\bea
\Psi(\tau \rightarrow \infty) & = &
 \lim_{\tau \rightarrow \infty} e^{-({\cal{H}}-E_T) \tau} \Psi_V 
  =  \lim_{\tau \rightarrow \infty} \sum_i \alpha_i e^{-(E_i - E_T) \tau} \Psi_i \nonumber \\
 & \longrightarrow & \alpha_0 e^{-(E_0-E_T) \tau} \Psi_0  ~,
\eea
where we have shifted the origin of energy to $E_T \approx E_0$ to control the norm
of $\Psi(\tau \rightarrow \infty)$. In practice, the time evolution is carried out in $n$ small steps
\beq
e^{-({\cal{H}}-E_T) \tau} = \prod e^{-({\cal{H}}-E_T)\Delta \tau}~,~~\Delta \tau  = \tau /n ~,
\label{eqn_slicing}
\eeq 
and $E_T$ is tuned to keep $\langle \Psi(\tau) | \Psi(\tau) \rangle$ constant.
The tuned $E_T$ provides the growth estimate of the true $E_0$. An alternative
method for calculating
 the ground state energy, often with smaller statistical error, is given by the mixed estimate (see Fig. \ref{fig_four})
\beq
\langle \mbox{$\mathcal{H}$} \rangle_{mix} = \frac{\langle \Psi_V| {\cal{H}} | \Psi(\tau \rightarrow \infty) \rangle}{\langle \Psi_V| \Psi(\tau \rightarrow \infty) \rangle}
 = E_0 \frac{\langle \Psi_V| \Psi(\tau \rightarrow \infty) \rangle}{\langle \Psi_V| \Psi(\tau \rightarrow \infty ) \rangle} = E_0~.
\eeq

In general, the time evolution operator or propagator is not known for arbitrary large value of $\tau$
except for few simple systems. However, we can obtain small time propagator
with controllable errors for any Hamiltonian with static potentials that depend only on the 
positions of the particles denoted by a $3N$-dimensional configuration vector 
${\bf{R}}= \{ {\bf{r}}_1, {\bf{r}}_2, \dots ; {\bf{r}}'_1, {\bf{r}}'_2, \dots \}$.
This is the motivation to write the time evolution as a product of many
short time operators (Eq. \ref{eqn_slicing}). We define the Green's function
\beq
G({\bf{R}},{\bf{R'}})  =  \langle {\bf{R}} | e^{-({\cal{H}}-E_T) \Delta \tau}| {\bf{R'}}\rangle~.
\eeq
The propagation equation becomes
\begin{equation}
\label{propeq}
\Psi({\bf R},\tau+\Delta \tau) = \int d{\bf R'} G({\bf R},{\bf R'}) \Psi({\bf R'}, \tau) \,.
\end{equation}
The primitive approximation to this Green's function is
\beq
G({\bf R},{\bf R'})  \approx 
e^{-(V({\bf R})- E_T)\frac{\Delta \tau}{2}} G_0({\bf R},{\bf R'}) 
e^{-(V({\bf R'})- E_T)\frac{\Delta \tau}{2}} ~,
\label{gst}
\eeq
where $V({\bf R}) = \sum\limits_{i,j'} v(r_{ij'})$ and
 $G_0({\bf R},{\bf R'})$ is the Green's function for $A$ free particles
\beq
G_0({\bf R},{\bf R'}) = \left [ \frac{m}{2\pi\hbar^2\Delta\tau}\right ]^{\frac{3}{2}A}
e^{ \left[ \frac{-m({\bf R}-{\bf R'})^2}{2 \hbar^2 \Delta \tau }\right ]}~.
\label{g_free}
\eeq
 This approximation has errors of order $\Delta \tau ^3$.
 The total error after $n$ time steps is of the order 
 $ \sim n \Delta \tau^3 = \frac{\tau^3}{n^2}$. The corrections
to this expression can be sampled to make an exact algorithm.
Here we use the more common method and make this error as small as we want, 
 by increasing the number of steps $n$.
 In practice, this error is made smaller than the statistical sampling errors of the Monte Carlo integration.

A naive quantum Monte Carlo algorithm could start with $N_s$ configuration vectors ${\bf{R}}_i$
sampled from $\left |\Psi_V \right |$. These provide the approximate representation
\begin{equation}
\Psi_V({\bf R}) \simeq \sum_{i=1}^{N_s} w_i\delta({\bf{R}}-{\bf{R}}_i)
\label{init}
\end{equation}
where $w_i=1$ or $-1$ depending on the sign of $\Psi_V$. The accuracy of this representation
increases with the number of samples $N_s$. Inserting Eq. \ref{init} into Eq. \ref{propeq} and using the short time approximation,
gives $\Psi({\bf{R}},\Delta \tau)$ as a sum of normalized gaussians times 
weight factors containing the product of the original $w_i$ and the 
exponentials in short time Green's function (Eq. \ref{gst}).
Sampling a position from each of the $N_s$ gaussians gives a representation of $\Psi({\bf{R}},\Delta \tau)$ as
a sum of delta functions times weight factors with signs. This process is repeated $n$ times to obtain $\Psi({\bf{R}},\tau= n\Delta \tau)$.
During the evolution, large magnitude weight factors
are converted into multiple copies while small factors are sampled
and kept with unit magnitude new weight with a probability proportional to the
magnitude of the old weight. The random walk of the weighted $\delta$-function 
samples representing the propagation of $\Psi({\bf{R}},\tau)$, 
therefore consists of diffusing and
branching and the number of samples at each time step can vary.

This algorithm suffers from the fermion sign problem.
For $N_s$ samples ${\bf{R}}_i$, $1 \leq i\leq N_s$, and weights $w_i$,
the denominator of a matrix element such as the mixed energy
will be the sum $\sum_{i=1}^{N_s} w_i \Psi_V({\bf{R}}_i)$. Each $w_i$
carries the sign of the initial sample from $\Psi_V$, and if the path
of the sample $i$ 
has crossed nodes of $\Psi_V$ odd number of times, the contribution
to the sum will be negative. For large times the contribution of these
negative paths almost completely cancel the contribution of positive
paths that have not crossed nodes or crossed an even number of times.
The signal dies out exponentially compared to the statistical
noise. The numerator suffers from the same problem.

The fixed node\cite{anderson75}  approximation deals
with the fermion sign problem by restricting the path so that
crossings of the nodal surface are not allowed.
When this constraint is imposed with the nodal surface of the exact 
fermion ground state, $\Psi({\bf R},\tau)$ converges to that state. Imposing
the nodal surface from an antisymmetric trial function gives an upper
bound
\beq
\lim_{\tau \rightarrow \infty} \langle {\cal{H}} \rangle_{mix,\tau} ~~ \ge ~~ E_0~.
\eeq
We impose the fixed-node constraint with the nodes of our trial
function $\Psi_V({\bf R})$.

Importance sampling is used to control the fluctuations of the weights.
The propagation equation is modified by multiplying by a positive importance
function. Since we are using the fixed node approximation, the paths have zero probability of crossing the nodes, and
we can take the importance function to be the absolute value of $\Psi_V$.
The propagation equation now becomes
\begin{equation}
\left [ \left |\Psi_V({\bf R})\right |\Psi({\bf R},\tau+\Delta \tau) \right ] =
\int d{\bf R'} \frac{\left |\Psi_V({\bf R})\right |}{\left |\Psi_V({\bf R'})\right |}G({\bf R},{\bf R'})
\left [ \left |\Psi_V({\bf R'}) \right |\Psi({\bf R'},\tau) \right ]\,.
\label{propeqimp}
\end{equation}

The short time approximation for the importance sampled Green's function is
\begin{eqnarray}
&~&\frac{\left |\Psi_V({\bf R})\right |}{\left |\Psi_V({\bf R'})\right |}G({\bf R},{\bf R'}) \nonumber \\ 
&=&
G_0\left ({\bf R},{\bf R'}+\frac{\hbar^2 \Delta \tau}{2m} \nabla \ln |\Psi_V({\bf R})|^2 \right)
\left \{\frac{\left |\Psi_V({\bf R})\right |G({\bf R},{\bf R'})}{\left |\Psi_V({\bf R'})\right |
G_0\left ({\bf R},{\bf R'}+\frac{\hbar^2 \Delta \tau}{2m} \nabla \ln |\Psi_V({\bf R})|^2 \right)
} \right \}
\nonumber\\
&\approx&
G_0\left ({\bf R},{\bf R'}+\frac{\hbar^2 \Delta \tau}{2m} \nabla \ln |\Psi_V({\bf R})|^2 \right)
\left \{
e^{-\left (\frac{1}{2}[E_L({\bf R})+E_L({\bf R'})]-E_T\right )\Delta \tau}
\right \}
\end{eqnarray}
where the local energy is
\begin{equation}
E_L({\bf R}) = \frac{{\cal{H}} \Psi_V({\bf R})}{\Psi_V({\bf R})}
\end{equation}
Since $G_0$ is still a normalized gaussian, the only changes to the
naive algorithm are the sampling of the drifted Gaussian, and the
new weight given by the terms in the braces. Notice that if $\Psi_V$
is close to the ground state of ${\cal{H}}$, $E_L({\bf R})$ will have less fluctuations
than $V({\bf R})$, and the branching of the walk is much reduced. Any paths that
cross a node due to the short time approximation are eliminated.

For $N_s$ samples ${\bf R}_i$, all with weight $w_i=1$, at time $\tau$, the mixed energy becomes the average of the local energy
\begin{equation}
\langle {\cal H}\rangle_{mix} = \frac{\sum\limits_{i=1}^{N_s} E_L({\bf R}_i)}{N_s} \,.
\end{equation}

Since the fixed node calculations give an upper bound to the ground state energy, our
strategy (see Ref. \cite{carlson2003}) is to choose a trial wave
function with variable nodal surfaces and minimize the fixed node GFMC $\langle {\cal{H}} \rangle_{mix}$.

The trial wave function $\Psi_V({\bf{R}})$ is now used in three different contexts; 1) as the
initial guess of the ground state, 2) as the importance function in Eq. \ref{propeqimp},
and 3) as the node restriction function. The nodes of the Jastrow-Slater wave function (Eq. \ref{locv_wf})
equal those of noninteracting Fermi gas and can not be varied. So that wave function is not
useful for present studies.

From physical considerations, a better trial wave function must reflect the fact that 
the fermions with attractive interaction can form bound Cooper pairs in the ground state. And from
 mathematical considerations, the trial wave function must have variable nodal surface, which can
be varied to minimize the fixed node GFMC energy. The BCS wave function is such a wave function.
 Commonly, we write
\bea
|BCS \rangle_\theta &=& \prod_p (u_{{\bf k}_p} + e^{i\theta}v_{{\bf k}_p} \hat{a}^{\dagger}_{{\bf k}_p \uparrow}
\hat{a}^{\dagger}_{{-\bf k}_p \downarrow})|0\rangle ~ ,  \label{bcs_wf0} \\ 
u_{{\bf k}_p}^2 + v_{{\bf k}_p}^2 &=& 1 ~, \nonumber 
\eea
where $|0 \rangle$ denotes the vacuum and $u_{{\bf k}_p}$ and $v_{{\bf k}_p}$ are real positive numbers.
However, this wave function does not correspond to a definite number of particles. 
In fact, expanding the wave function we can write
\beq
|BCS \rangle_\theta = |0\rangle ~ + e^{i\theta}\hat{P}^{\dagger} |0\rangle ~ +  e^{i 2 \theta}(\hat{P}^{\dagger})^2|0\rangle ~ +
  e^{i 3 \theta}(\hat{P}^{\dagger})^3|0\rangle ~ + \dots ~.
\label{bcs_exp}
\eeq
where $\hat{P}^{\dagger} = \sum\limits_p \frac{v_{{\bf k}_p}}{u_{{\bf k}_p}} \hat{a}^{\dagger}_{{\bf k}_p \uparrow}
\hat{a}^{\dagger}_{{-\bf k}_p \downarrow}$ is the pair creation operator. The component that
corresponds to $A$ particles or $M = A/2$ pairs can be obtained by transforming
\bea
|BCS \rangle_A &=& \frac{1}{2\pi} \int\limits_0^{2\pi}  e^{-i \theta M}~|BCS \rangle_\theta ~ d\theta ~,\nonumber \\
& = & (\hat{P}^{\dagger})^{M} |0\rangle ~.
\eea
This component can be written as an antisymmetrized product of the pair wave functions $\phi(r_{ij'})$
\bea
\Psi_{BCS}({\bf{R}}) &=& {{\cal A}}[\phi(r_{11'}) \phi(r_{22'}) ... \phi(r_{MM'})] ~, \\
\phi(r) &=&\sum\limits_p \frac{v_{{\bf k}_p}}{u_{{\bf k}_p}} e^{i{\bf k}_p\cdot{\bf r}} =\sum_p \alpha_p e^{i{\bf k}_p\cdot{\bf r}} ~, 
\nonumber
\eea
where the number of up spin particles ($M$) is equal to the number of down spin particles ($M'$).
 The variational parameters $\alpha_p$ are real positive numbers. 
The free fermion gas, Slater wave function is just a particular case of this wave function 
when $\alpha_p \neq 0$ for $|{\bf k}_p| \le k_F$ and $=0$ for $ |{\bf k}_p| > k_F$.

 We also consider systems having unpaired particles.
In particular, we can have $M$ pairs and 1 unpaired up or down spin particle.
This generalization is necessary as the gap energy $\Delta$ is calculated from the odd-even staggering of the
ground state energy \cite{carlson2003}. With 1 unpaired ($\uparrow$ or $\downarrow$ spin) particle in the state 
$\psi_{{\bf{k}}_{u}}({\bf r})$, with momentum ${\bf{k}}_{u}$, the trial wave function is given by \cite{bouchaud1988}
\beq
\Psi_{\rm BCS}({\bf{R}}) =
{{\cal A}} \left \{ 
\left[ \phi(r_{11'})...\phi(r_{MM'})\right] \psi_{{\bf{k}}_{u}}({\bf r}) \right \} ~.
\label{bcs_odd}
\eeq
The ground state is expected to have $|{\bf{k}}_{u}| \approx |{\bf{k}}_{F}|$ in the weakly interacting regime and
${\bf{k}}_{u} \rightarrow {\bf{0}}$ in the strongly interacting regime.
This wave function can be calculated as a determinant \cite{bouchaud1988},\cite{carlson2003}, which makes
the numerical calculations relatively simple.

 Quantum Monte Carlo calculations use 
a finite number of particles in a cubic periodic box of volume $L^3$ to simulate the infinite uniform system. The momentum
vectors in this box are discrete
\beq
{\bf k}_p = \frac{2 \pi}{L} ( n_{px} \hat{x} + n_{py} \hat{y} + n_{pz} \hat{z}) ~,
\eeq
 and the system has a shell structure with closures occurring when the total number of particles = 2, 14, 38, 54, $\dots$ for spin-$1/2$ fermions.
The shell number $I$ is defined such that $ I = n_x^2 +n_y^2 + n_z^2$, and $E_I = \frac{\hbar^2}{2m} \frac{4 \pi^2}{L^2} I$.

In the present calculations, the pair wave function $\phi({\bf{r}})$ has the assumed form
\bea
\phi({\bf r}) &=&\tilde{\beta} (r) + \sum_{p,~I \leq I_C} \alpha_I e^{ i {\bf{k}}_p \cdot {\bf r}}~, \\
\tilde{\beta}(r) &=& \beta(r)+\beta(L-r)-2 \beta(L/2)~~~ \mbox{for}~~ r \le L/2~, \nonumber  \\
 &=& 0 ~~~~~~~~~~~~~~~~~~~~~~~~~~~~~~~~~~~~~ \mbox{for} ~~ r > L/2~, \nonumber \\
\beta (r) &=& [ 1 + \gamma b  r ]\ [ 1 - e^{ - c b r }] \frac{e^{ - b r }}{c b r}~.  \nonumber
\eea
Here $I_c = 4$ is a cut off shell number. We assume that the contributions of shells with 
$ I > I_c$ to the pair wave function can be approximated by a spherically symmetric 
function $\tilde{\beta}(r)$ of range $L/2$. We further reduce the
statistical fluctuations by using the Jastrow factor along with $\Psi_{BCS}$ in the 
variational wave function:
\beq
\Psi_V({\bf R}) = \prod\limits_{i,j'} f(r_{ij'}) \Psi_{BCS}({\bf R})~.
\eeq
The Jastrow factor
does not change the nodal structure. Thus, the average value of the estimated energy is independent of
$f(r)$, but the statistical error is reduced by using the $f(r)$ from LOCV calculations.

It is convenient to require that $\frac{\partial \tilde{\beta}}{\partial r} = 0$ at $r=0$. 
This is because the local energy has terms like $\frac{1}{r}\frac{\partial \tilde{\beta}}{\partial r}$
which can have large fluctuations at the origin when $\frac{\partial \tilde{\beta}}{\partial r} \neq 0$ at $r=0$.
The factor $[1- e^{-cbr} ]$ cuts off $\frac{1}{cbr}$ dependence of $\beta$ at $br < \frac{1}{c}$.
The energies are not too sensitive to the parameter $c$, and its value is fixed at $10$. In addition,
$\gamma$ is chosen such that $\frac{\partial \tilde{\beta}}{\partial r} = 0$ at $r=0$; its value
is $6$ in the limit $L \rightarrow \infty$.

The variational parameters are $\{ \alpha_0, \alpha_1, \dots, \alpha_{I_c} \}$ and $b$.
We wish to find a set of these parameters that minimize the fixed node GFMC estimate of energy. 
 However, considering that we have to allow simultaneous variation of all the parameters, methods based on
 unguided variation become difficult, if not infeasible. Again, we rely on the
 GFMC procedure itself to optimize these parameters. Initial configurations are obtained with a
 random distribution of the parameters centered around a reasonable guess. Each of them is propagated
 according to the nodal constraints provided by their parameters with a single $E_T$. The paths with the
 smallest $ \langle {\cal{H}} \rangle_{mix}$ acquire large amplitudes or weights as $\tau \rightarrow \infty$.
 The average among these paths gives an optimization over the initial random distribution.
  This process is repeated several times until convergence is achieved.

When we have an odd number of particles, the ground state momentum ${\bf{k}}_u$ (Eq. \ref{bcs_odd}) is an additional
variational  parameter. We minimize the fixed node GFMC energy of systems with odd $A$ by
varying ${\bf{k}}_u$. As discussed in the results section, 
the magnitude of ${\bf{k}}_u$ changes from $k_F$ to $0$ as the interaction strength increases and we go from the 
weakly interacting BCS to strongly interacting BEC regime.
The gap energy is obtained from the odd-even staggering of the total energy
\beq
\Delta(2M + 1)=E(2M+1)-\frac{1}{2}(E(2M)+E(2M+2))
\label{exp_gap}
\eeq
In doing so, the effects of interaction among quasiparticles are neglected. 

\begin{figure}
\includegraphics[width=\columnwidth]{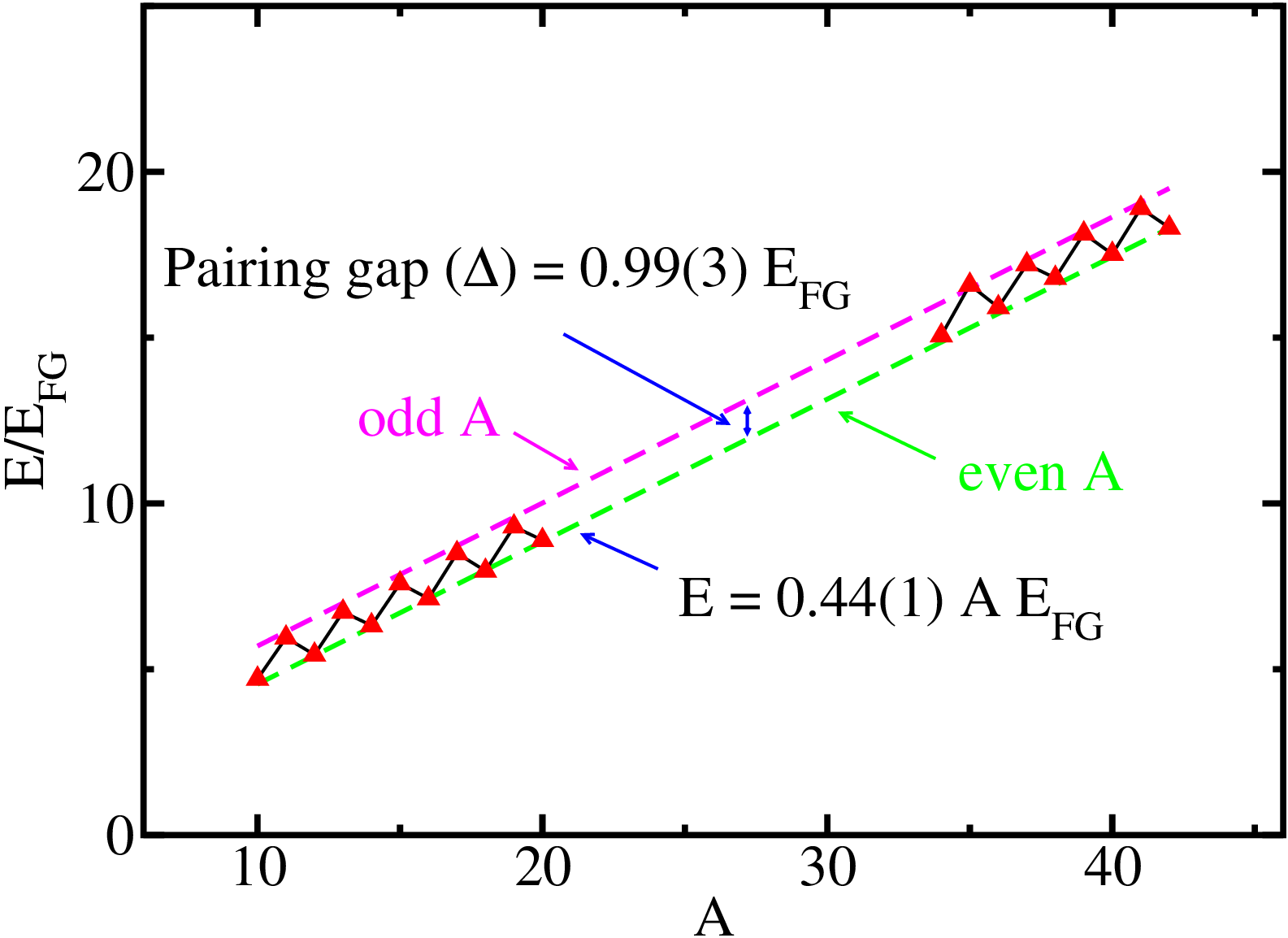}
\caption{$E(A)$ when $a k_F = -\infty$ from Ref. \cite{carlson2003}.}
\label{fig_five}
\end{figure}

Results for $a k_F =\infty$ are shown in Fig. \ref{fig_five}. The energy per particle $E/A$ and the gap $\Delta$
do not have a significant $A$ dependence in this case.
These results were reported in Ref. \cite{carlson2003}, and results for other values of $a k_F$ are presented in the next section.

\section{RESULTS}\label{sec4}
 The values of the parameters, $\alpha_{0-4}$ and $b$, of the BCS wave
function are to be determined by minimizing the fixed node GFMC mixed energy
for each value of $ak_F$ and $A$.  The minimum energy obtained is our estimate
for the ground state energy of the system.  The values of the parameters
that minimize this energy are not very sensitive to $A$, the number of particles in the box.
 We find it sufficient to determine the optimum parameters at $A=10,~14$ and $20$, and interpolate their
values in the $A=10,14$ and $A=14,20$ ranges.  The values of the parameters at these values of $A$ 
are listed in Table \ref{table_param}.

At $ak_F=-1$ the lowest energies are obtained without any short range
$\tilde{\beta}(r)$ and the optimum pair function has contributions only
from the states with $I \leq 3$.  This is consistant with the weak coupling BCS 
theory in which $\alpha_k$ goes to zero when $k-k_F$ becomes large. 

When $1/ak_F > -1$, lower energies are obtained with 
$\tilde{\beta}(r) \neq 0$. In most cases, the values of the
parameters do not change significantly between $A = 14$ and $20$.  
The values listed in Table \ref{table_param} for $A = 14$ are used for $14 \leq A \leq 20$.

 In the $0 < 1/ak_F \le 2$ range, the optimum values of the parameters of
$\Psi_{BCS}$ do not seem to change significantly with the $ak_F$.
We have not obtained any significant improvements to the energy
from varying the parameters in the region $1/ak_F > 0$.
In this region we retain the values found for $1/ak_F = 0$ and $A = 14$. 
Recall that only the nodal surfaces of the ground state wave function are
constrained by those of $\Psi_{BCS}$.  The complete $\Psi_V$ has an
additional product of Jastrow pair correlation functions $f(r_{ij'})$ which depends on $ak_F$, and
the true ground state wave function changes continuously with $ak_F$.

 The magnitude of the momentum ${\bf{k}}_u$ of the unpaired particle in the ground
state is also determined by minimizing the GFMC fixed node energy.  The minimum
values are listed in Table \ref{table_ku}.  In the weak coupling limit, the BCS ground
state for odd $A$ has $|{\bf{k}}_u| = |{\bf{k}}_F|$.  In the periodic box, the value of $k_F^2$ is
1 for $2 < A \le 14$, and 2 for $14 < A \le 38$ in units of $(2\pi/L)^2$.  
In the $ak_F = 0$ to $-3$ range, the minimum values of $k_u^2$ are as indicated
by the weak coupling BCS theory. However, in the $-10$ to $3$ range the $k_u^2$ is 1
for the entire range (11 to 19) of odd values of $A$ considered.  At $ak_F = 2$
the states with $k_u^2 = 0$ and 1 are almost degenerate, and for $1/ak_F > 0.5$
the ground states of odd $A$ systems have ${\bf{k}}_u = \bf{0}$, as expected when the
system consists of bound molecules condensed in the zero momentum state,
and the unpaired particle also in the ${\bf{k}}_u = \bf{0}$ state.

\begin{figure}
\includegraphics[width=\columnwidth]{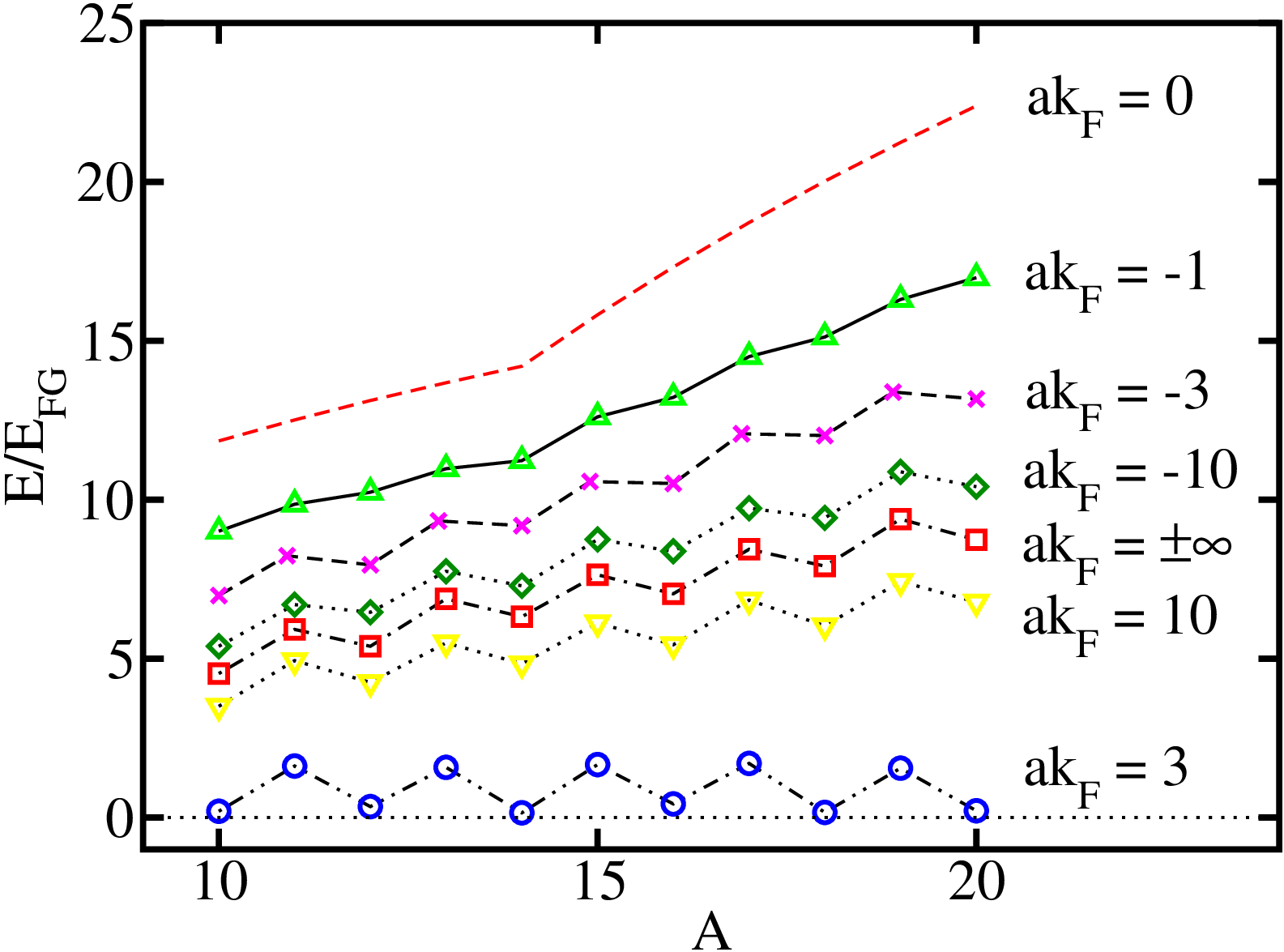}
\caption{ $E(A)$ for $1/a k_F \le 1/3$.}
\label{fig_six}
\end{figure}

\begin{figure}
\includegraphics[width=\columnwidth]{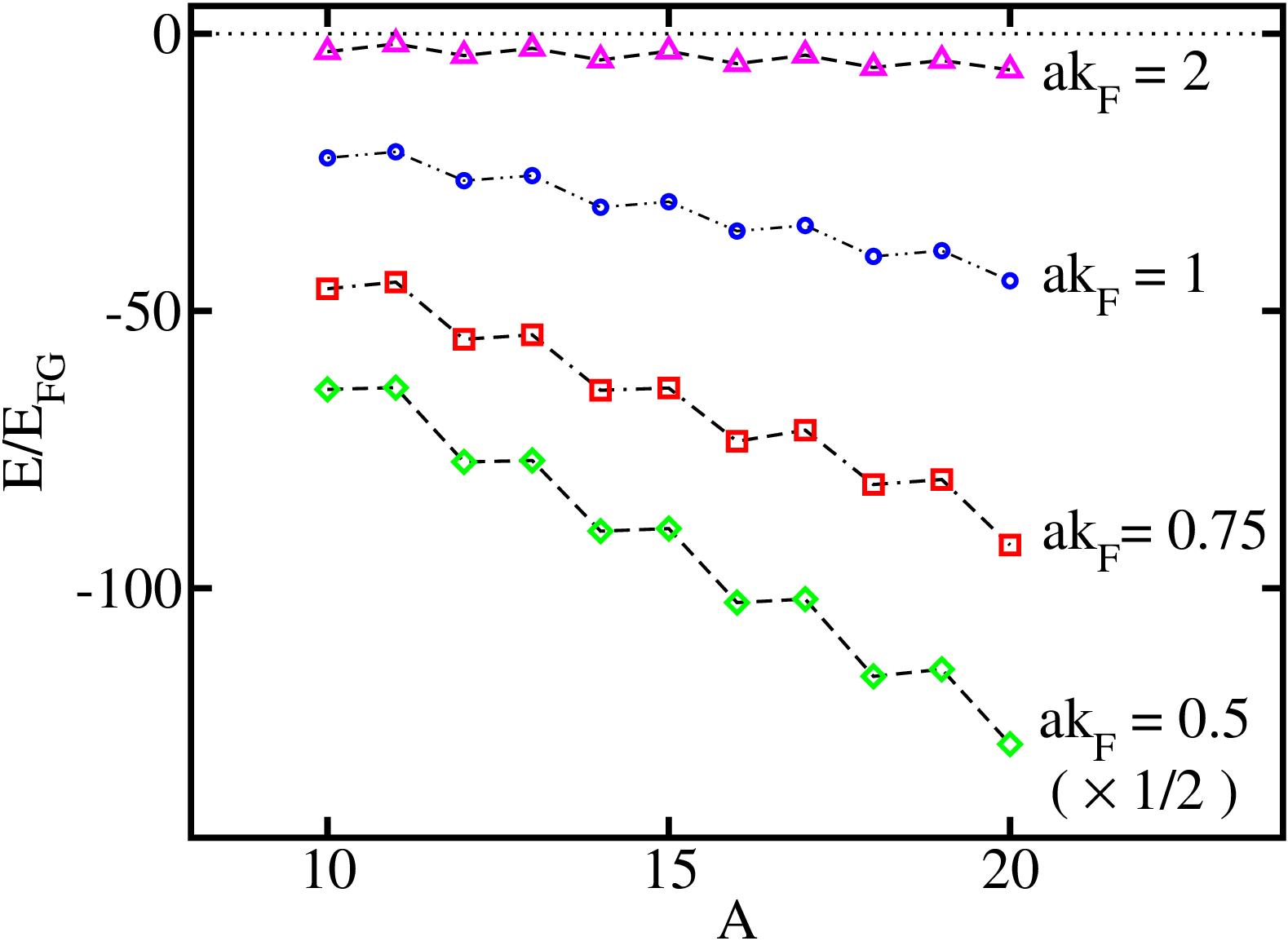}
\caption{$E(A)$ for $1/a k_F > 1/3$. The results for $ak_F = 0.5$ have been multiplied by 0.5 for graphing.}
\label{fig_seven}
\end{figure}

\begin{figure}
\includegraphics[width=\columnwidth]{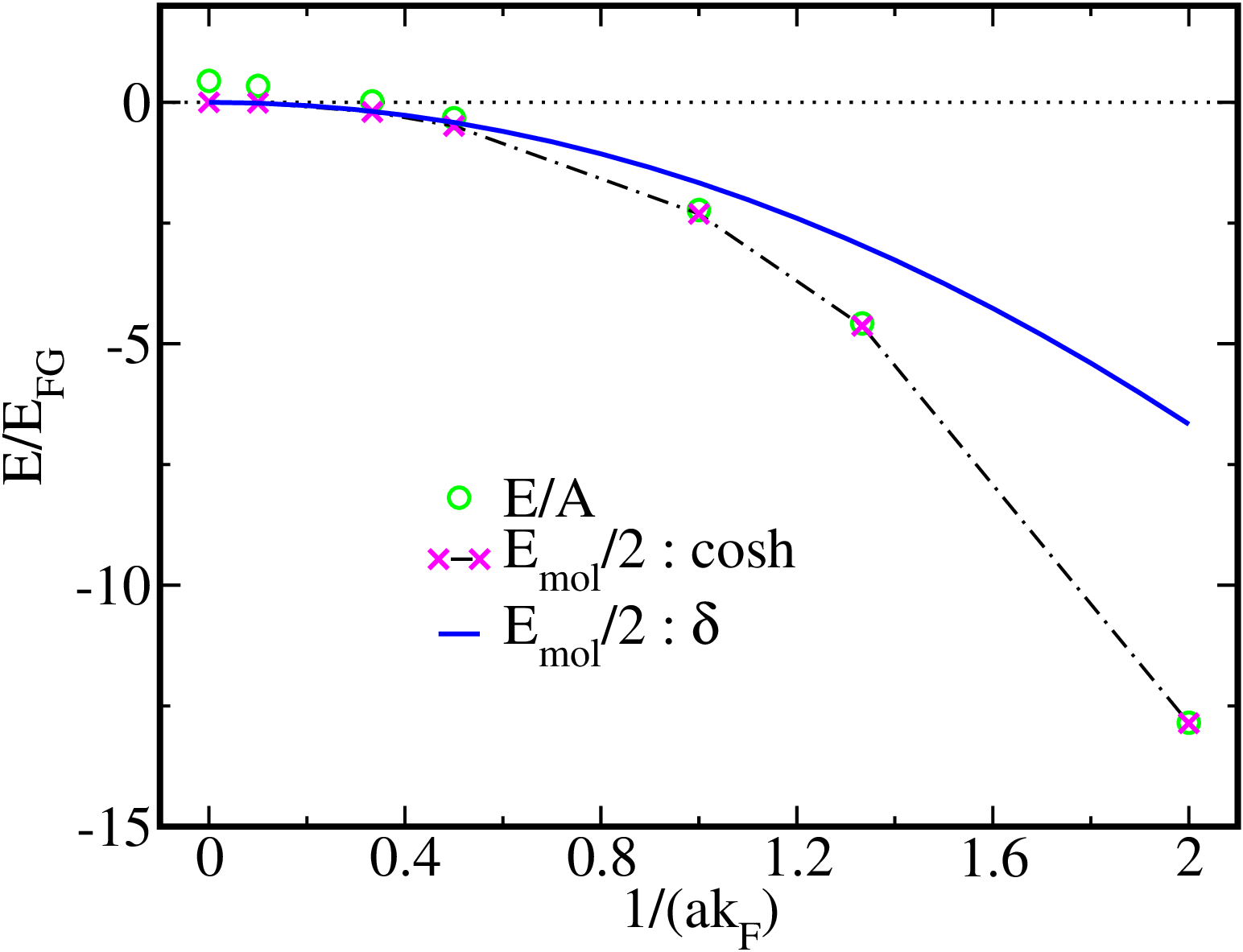}
\caption{$E/A$ and $E_{mol}/2$ for positive values of $1/ak_F$ for the $\cosh$ ($\mu r_0 = 12$) potential
and $E_{mol}/2$ for the delta-function potential.}
\label{fig_eight}
\end{figure}

   The calculated values of the ground state energy are shown in figures
\ref{fig_six} and \ref{fig_seven}.  The systems with $1/ak_F \leq 1/3 $ seem to have $E > 0$,
while those with $1/ak_F > 1/3$ can have $E < 0$.  When $1/a > 0$ the
two-body interaction is strong enough to bind two particles and form
molecules with energy $E_{mol}$.  The energy per particle, $E/A$
of the superfluid Fermi gas is compared with $E_{mol}/2$ in figure \ref{fig_eight}.  Within
the computational errors $E/A > E_{mol}/2$ (see Table \ref{table_strong}), however
at $1/ak_F \geq 0.5$ we find that $E/A$ is very close to $E_{mol}/2$.  
This behavior also indicates that at these values of $ak_F$ the system approaches that composed
 of Bose molecules forming a BEC. It has been argued that the interaction between these
molecules is weakly repulsive, with a molecule-molecule scattering length
given by $a_{mm} = 0.6 a$ \cite{petrov03}.  In this case the $E/A$ will
always be greater than $E_{mol}/2$, and the gas will have positive pressure,
$E/A$ increasing with the gas density or $k_F$.  

\begin{figure}
\includegraphics[width=\columnwidth]{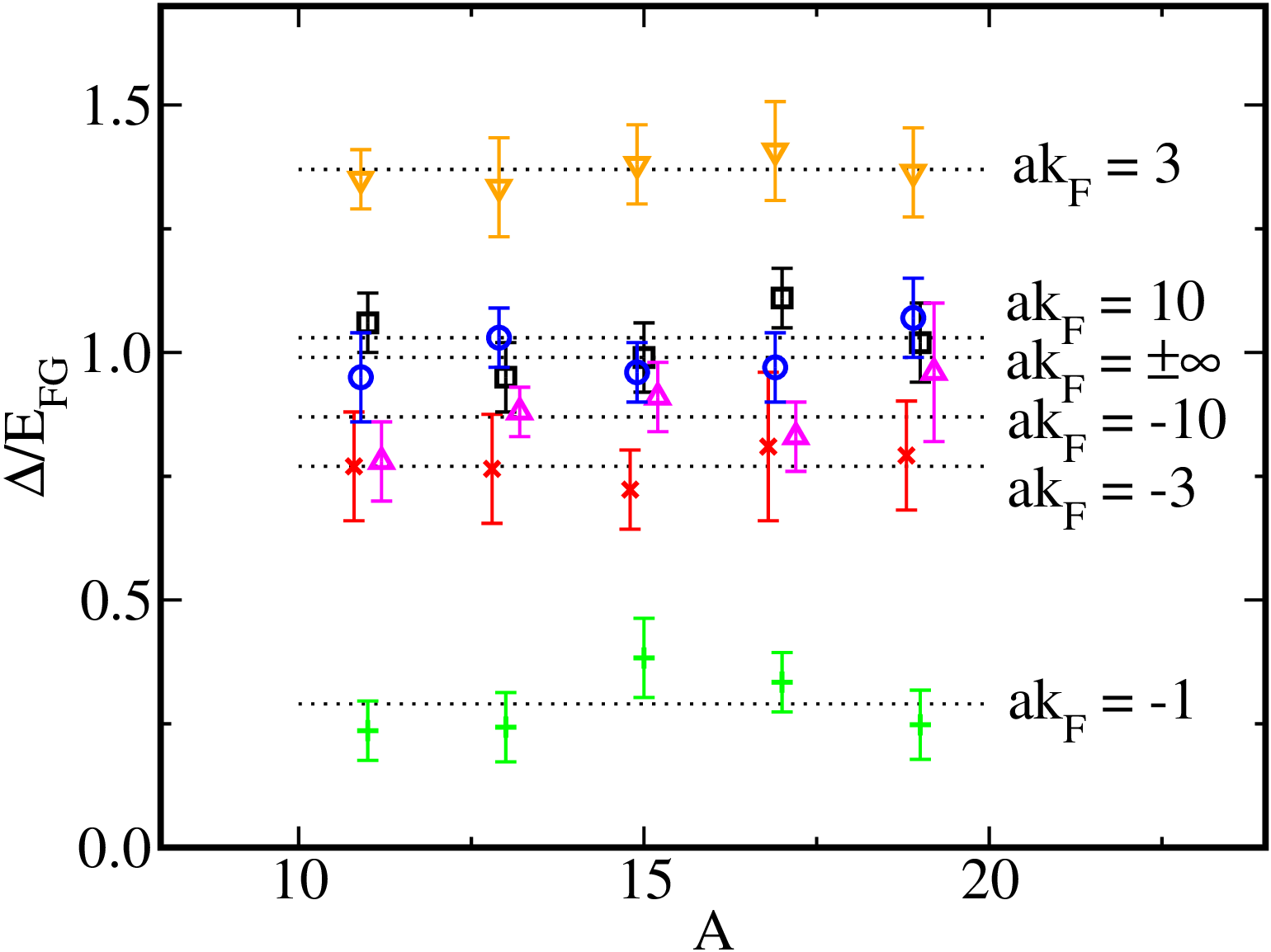}
\caption{$\Delta(A)$ for $1/a k_F \le 1/3$.}
\label{fig_nine}
\end{figure}

\begin{figure}
\includegraphics[width=\columnwidth]{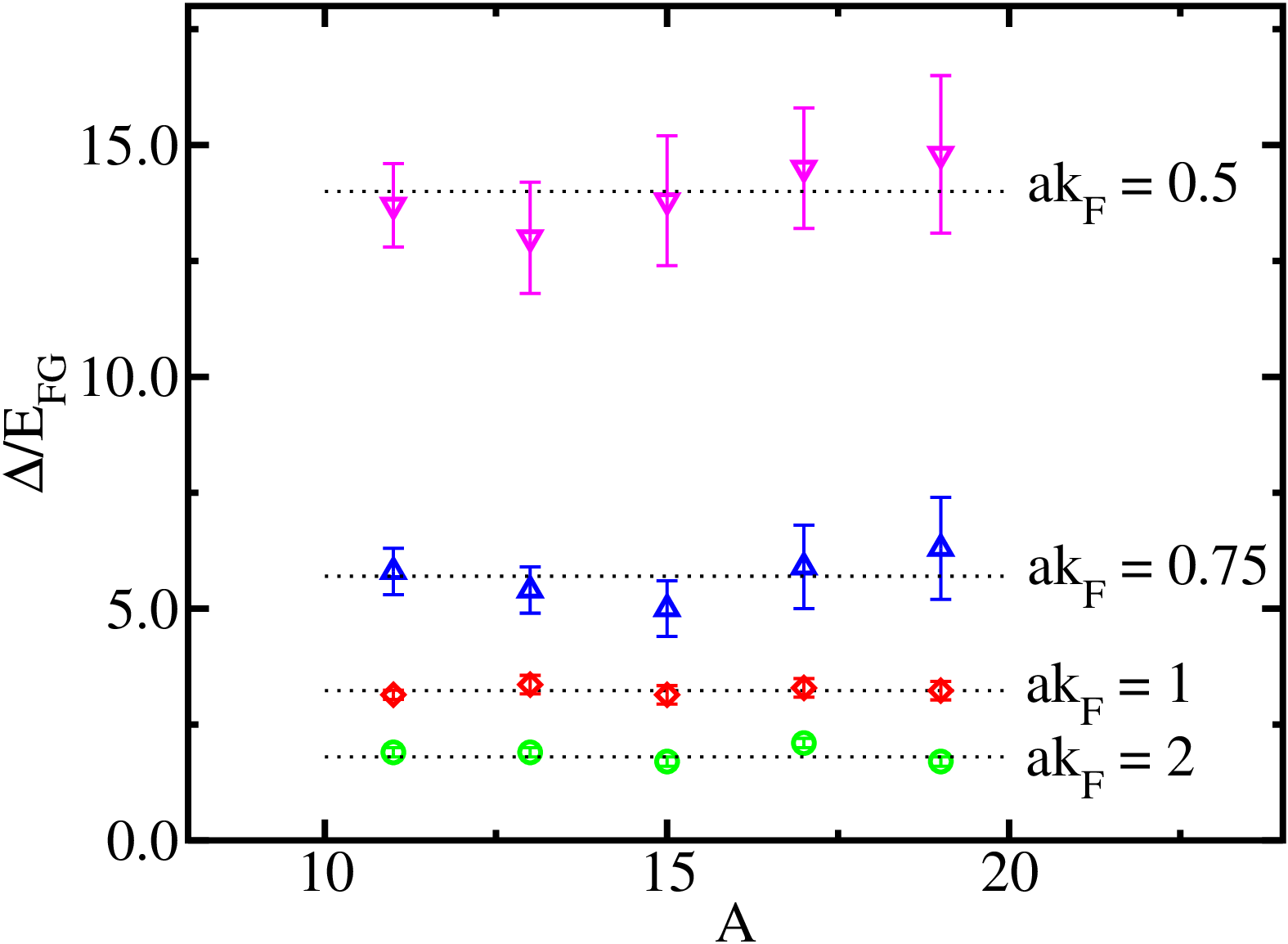}
\caption{$\Delta(A)$ for $1/a k_F > 1/3$.}
\label{fig_ten}
\end{figure}

 The pairing gaps calculated from the odd-even energy difference (Eq. \ref{exp_gap})
are shown in figures \ref{fig_nine} and \ref{fig_ten}.  These gaps are not very sensitive to $A$,
and they are compared with the predictions of BCS \cite{leggett1980} and Gorkov \cite{gorkov61}
estimates given by
\bea
\Delta_{BCS} & = & \frac{8}{e^2} ~ T_F ~ e^{\pi/(2ak_F)} ~, \nonumber \\
\Delta_{Gorkov}& = & \left( \frac{2}{e} \right)^{7/3} T_F ~ e^{\pi/(2ak_F)}~ 
= \frac{1}{2} \left( \frac{2}{e}\right)^{1/3} \Delta_{BCS}~.
\label{gap_old}
\eea
where the chemical potential is approximated by $T_F$ as when $1/ak_F << 0$.
At $1/ak_F << 0$ the calculated gaps are in between these estimates, while
at positive values of $1/ak_F$ they approach $-E_{mol}/2$ as expected for a gas
of Bose molecules (see Fig. \ref{fig_ten} and Table \ref{table_strong}).

\begin{figure}
\includegraphics[width=\columnwidth]{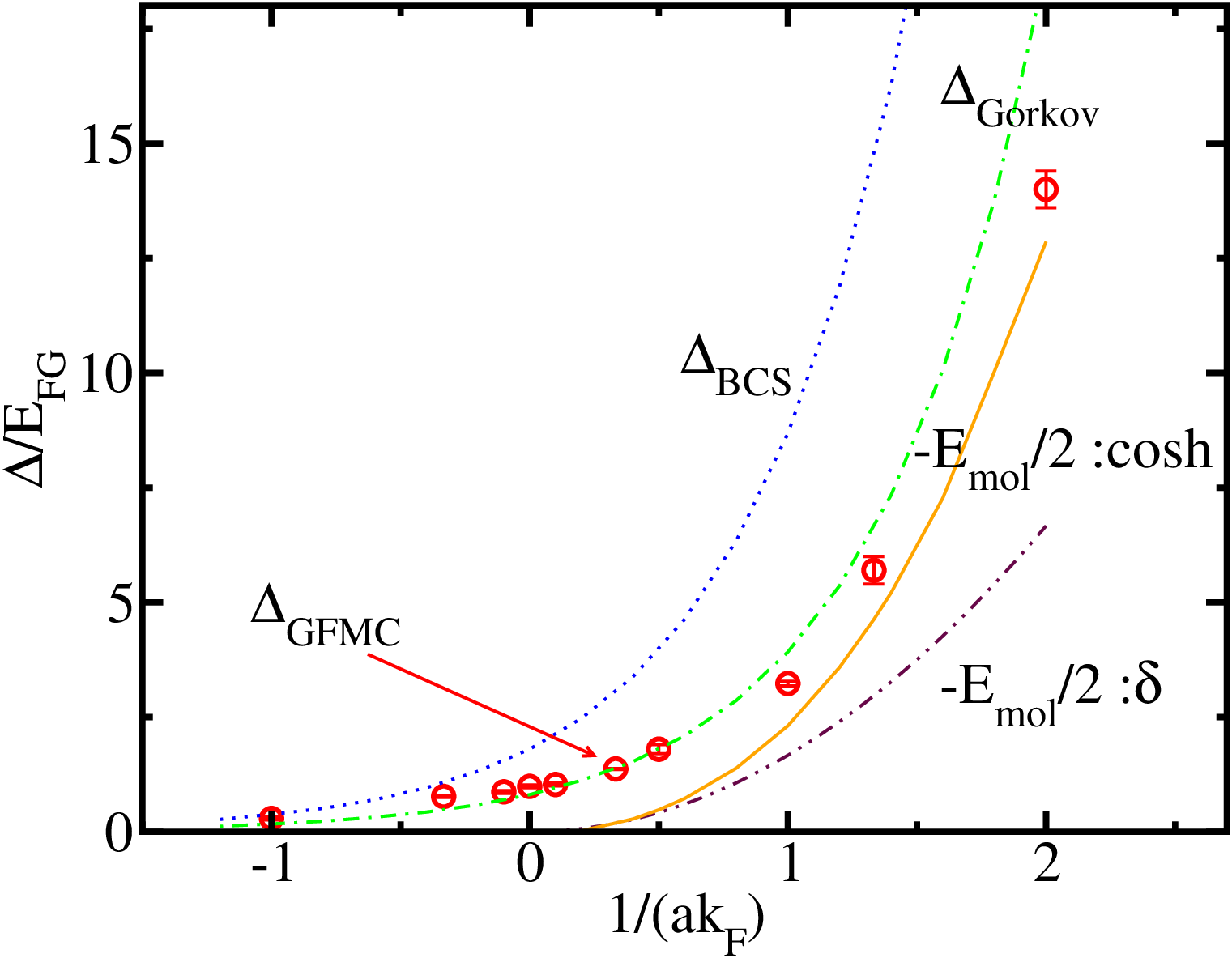}
\caption{Calculated values of $\Delta_{GFMC}(ak_F)$ ($\cosh$, $\mu r_0 = 12$ potential) are compared with various estimates of
 $\Delta(ak_F)$ and $-E_{mol}/2$. The BCS and Gorkov estimates do not depend on the shape of the potential, while
$-E_{mol}/2$ is shown for both $\cosh$ (solid line) and delta-function (dash double dot) potentials.
$\Delta_{BCS}$ and $\Delta_{Gorkov}$ assume the chemical potential $\approx T_F$ throughout the whole
range of $ak_F$(see equations \ref{gap_old}).}
\label{fig_eleven}
\end{figure}

   Figures \ref{fig_eight} and \ref{fig_eleven} also show the $E_{mol}$ for a delta-function interaction
in addition to those for the present $\cosh$ potential with $\mu r_o = 12$.  The
two potentials give essentially the same results for $1/ak_F < 0.5$, but at
larger values the $\cosh$ potential is more attractive.  The values of the $rms$
radius $R_{rms}$ of the molecule are listed in Table \ref{table_strong}.  At large values of $1/ak_F$
the $\mu R_{rms}$ is not very large for the present choice of $\mu r_0=12$, and much
larger values of $\mu$ should be used to approximate the delta-function
interaction.

\begin{figure}
\includegraphics[width=\columnwidth]{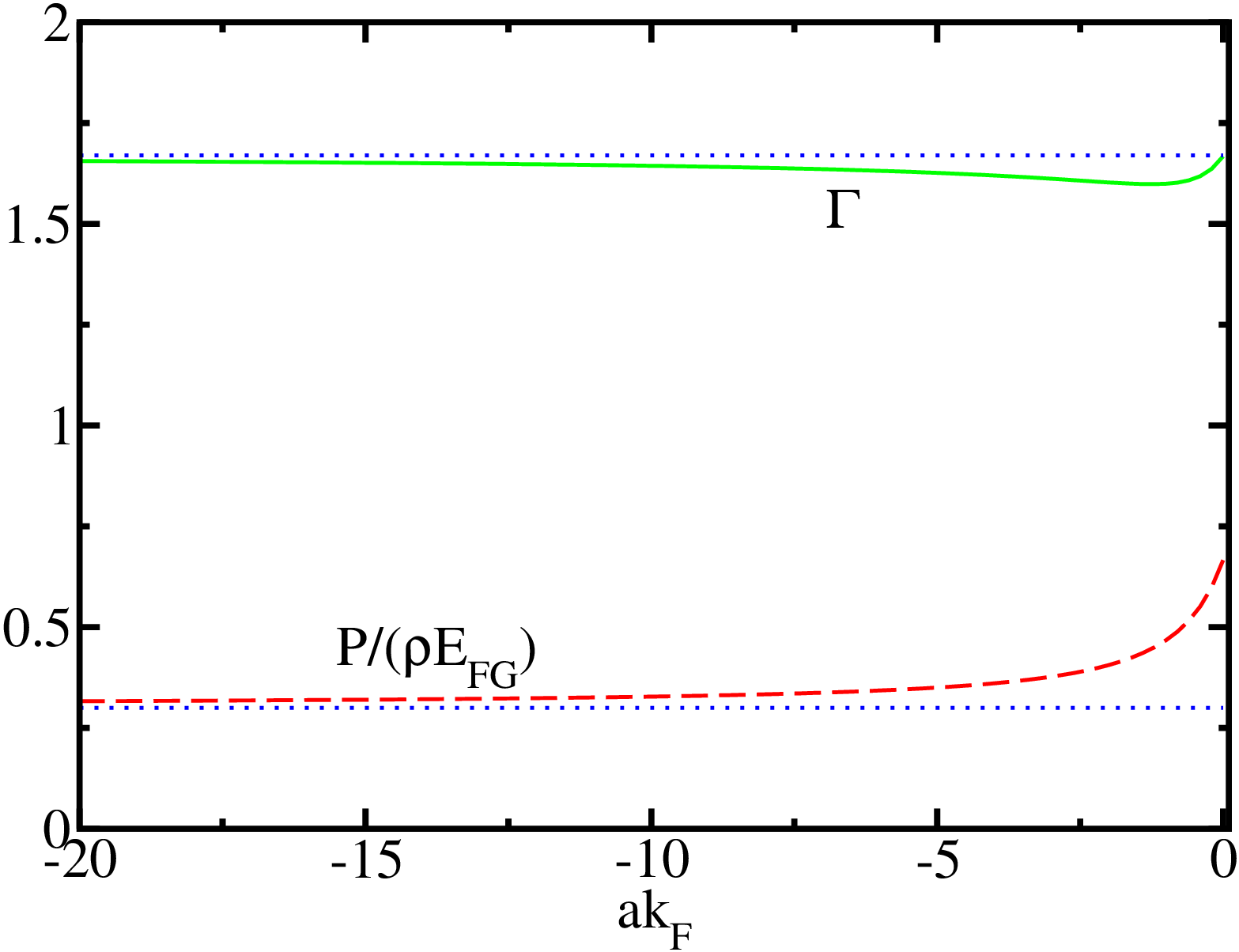}
\caption{Pressure($P$, dashed curve) and adiabatic index($\Gamma$, continuous curve)
 in the BCS regime ($ak_F < 0$). In the dilute limit ($ ak_F \rightarrow 0$) we have $P/(\rho E_{FG}) \rightarrow \frac{2}{3}$ and
 $\Gamma \rightarrow \frac{5}{3}$. In the dense limit ($ ak_F \rightarrow -\infty$) we have $P/(\rho E_{FG}) \rightarrow 0.44\frac{2}{3}$ and
 $\Gamma \rightarrow \frac{5}{3}$. $\Gamma$ has a minimum at $ak_F \sim -1.3$.}
\label{fig_twelve}
\end{figure}

 The pressure ($P = \rho^2 \frac{\partial(E/A)}{\partial \rho}$) and the adiabatic index ($\Gamma = \frac{\rho}{P}\frac{\partial P}{\partial \rho}$) of the
 superfluid gas in the range $-20 < ak_F < 0$ are shown in Fig. \ref{fig_twelve}. For noninteracting Fermi
 gas ($ak_F = 0$), $E/A = E_{FG}$ and we have $P = \frac{2}{3}\rho E_{FG}$ and $\Gamma = \frac{5}{3}$. In the limit 
 $ak_F \rightarrow 0$ we can use the low density expansion (Eq. \ref{low_exp}) to obtain
 \bea
 P(ak_F \rightarrow 0) & \approx & \frac{2}{3} \rho E_{FG} (1 + \frac{5}{3 \pi} ak_F ),
 \nonumber \\
 \Gamma(ak_F \rightarrow 0) & \approx & \frac{5}{3} + \frac{5}{9\pi} ak_F.
 \eea
  In the $ak_F \rightarrow -\infty$ limit, we have $E/A = \xi E_{FG}$, therefore 
 \bea
 P(ak_F \rightarrow -\infty) & = & \frac{2}{3}\xi\rho E_{FG} \nonumber  \\
\Gamma(ak_F \rightarrow -\infty) & = & \frac{5}{3} 
 \eea
 where $\xi = 0.44 \pm 0.01$ according to the present calculations. 

 The calculated value of $P(ak_F)/ (\rho E_{FG})$ is $\frac{2}{3}$ at $ak_F = 0$ and decreases monotonously to $\frac{2}{3} \xi$ as
 $ak_F \rightarrow -\infty$. However, The adiabatic index $\Gamma(ak_F)$, is $\frac{5}{3}$ for both $ak_F = 0$ and $ak_F = -\infty$,
 and has a minimum value of $\sim 1.6 $ at $ak_F \sim -1.3$.
  
\section{CONCLUSIONS}\label{sec5}
 The present work shows that accurate calculations of the pairing gaps and
 energies of superfluid Fermi gases are possible with the fixed node GFMC method. The unknown nodal surfaces
 can be determined variationally by minimizing the 
 fixed node GFMC energy. This method gives the exact result in the
 $1/ak_F \rightarrow -\infty$ (Fermi gas) and $1/ak_F \rightarrow +\infty$ 
 (BEC of molecules) limits for short range attractive
 interaction, and seems to overcome the fermion sign problem. An alternative method based on path
 integral Monte Carlo simulations is also being developed \cite{shumway00}.

 Our results are in qualitative agreement with the known BCS-BEC crossover model (see Leggett \cite{leggett1980})
 where gap and chemical potential($\mu_0$) are calculated self consistently.  
 The gap is determined as the minimum of the Bogoliubov quasiparticle energy
 $E_{\bf k} = \sqrt{(\epsilon_{\bf k} - \mu_0 )^2 + |\Delta'_{\bf k}|^{2}}$, where $\epsilon_{\bf k} = \frac{\hbar^2 k^2}{2m}$
 is the single particle excitation energy and $\Delta'_{\bf k}$ is the gap parameter.
  Two limiting cases were considered in this referenced article. For $1/ak_F \rightarrow -\infty$, 
   $\mu_0 \approx T_F > 0$, the minimum of $E_{\bf k}$ occurs at $k = k_F$ and the minimum quasiparticle energy
 $\Delta$ = $\Delta'_{k_F} = \frac{8}{e^2}~T_F~e^{\pi/(2ak_F)}$. 
 However, for $1/ak_F \rightarrow +\infty$, $\mu_0 \approx E_{mol}/2 < 0$,
 the minimum of $E_{\bf k}$ is at $k = 0$, and its value $\Delta = |\mu_0| \sim |E_{mol}|/2$ because $\Delta'_{\bf k} \sim 0$. 
  The BCS-BEC crossover takes place when $\mu_0 =0$ and this corresponds to $ak_F$ positive and of the order 1.
  The odd-even staggering $\Delta(2M+1)$ given by Eq. \ref{exp_gap} presumably equals the minimum quasiparticle energy in 
  the limit $M \rightarrow \infty$. 
 
 According to Leggett's description, in the weak BCS superfluids the ground state of systems 
 with odd number of particles is expected to have momentum $k_F$, while in the molecular liquid with BEC it is
 expected to have zero momentum. With this criteria the
 calculated values of $k_{u}^2$ (Table \ref{table_ku}) suggest that the BCS to BEC transition occurs in the
 range $-0.5 < 1/ak_F < 0.5$. It appears to be a smooth transition or crossover. 

  A recent experiment by Bartenstein et al. also seems to corroborate some of our findings.
 In fact, in their paper \cite{bartenstein2004} BCS-BEC crossover regime for $^6$Li is reported to be $-0.5 \alt 1/ak_F \alt 0.5$. 
 In addition, in the unitary limit ($ak_F = \pm\infty$) they measured $E/A = 0.32^{+13}_{-10} E_{FG}$ which 
 includes within its range our result $E/A = (0.44 \pm 0.01) E_{FG}$. 
 
  We can notice that in the BCS regime $\Delta$ is
 much smaller than $E/A$, while in the BEC regime $\Delta \sim |E|/A \sim |E_{mol}|/2$. However, in the
 transition region $\Delta$ is significantly larger than $|E|/A$.
 
\section{ACKNOWLEDGEMENTS}\label{sec6}
 The work of J.C. is supported by the U. S. Department of energy under contract 
 No. W-7405-ENG-36, while that of S.Y.C. and V.R.P is partly supported by U.S. National Science Foundation via
  Grant No. PHY-00-98353.
 
\clearpage

\clearpage

\clearpage

\clearpage

\clearpage

\clearpage

\clearpage

\clearpage

\clearpage

\clearpage

\clearpage

\clearpage

\clearpage

\begin{center}
\begin{table}
\caption{Optimum values of the variational parameters.}
\begin{tabular}{cc|cccccc} 
\hline
~~$ak_F$~ &~ $A$~~ & ~~$\alpha_0$~~ & ~~$\alpha_1$~~& ~~$\alpha_2$~~& ~~$\alpha_3$~~& ~~$\alpha_4$~~& ~~$b$~~\\
\hline
\hline
-1&10& 1.00 & 0.05  & 0   & 0 & 0&NA\\
  &14& 1.00 & 1.00 & 0.010 & 0   &  0&NA \\
  &20& 1.00 & 1.00 & 0.104 & 0.024&  0&NA\\
\hline
-3&10 & 0.40 &0.165& 0.019&0.009 &0.002 & 1.13\\
  &14 & 0.28 & 0.280 &0.020&0.006 & 0.003 &1.05\\
\hline
-10&10 & 0.295 & 0.096 & 0 018 & 0.007 & 0.002 & 0.48 \\
   &14 & 0.220 & 0.130 & 0.019 & 0.007 & 0.003 & 0.44 \\
\hline
$\infty$ &10 & 0.315 & 0.103 & 0.020 & 0.010 & 0.003 & 0.50 \\
         &14 & 0.181 & 0.102 & 0.024 & 0.006 & 0.004 & 0.44 \\
\hline

\hline
\end{tabular}
\label{table_param}
\end{table}

\begin{table}
\caption{Values of $k_u^2$ in units of $(\frac{2\pi}{L})^2$.}
\begin{tabular}{c||c|c}
\hline
$ak_F$ & N=11,13 & N=15,17,19 \\
\hline
0   &  1 & 2 \\
-1  &  1 & 2 \\
-3  &  1 & 2\\
-10 &  1 & 1\\ 
$\infty$ & 1 & 1 \\
10  &   1 & 1\\
3   &   1 & 1\\
2   &  0 or 1 & 0 or 1 \\
1   &   0 & 0 \\
0.75 &  0 & 0 \\
0.5   &  0 & 0 \\
\hline
\end{tabular}
\label{table_ku}
\end{table}
\end{center}
\clearpage

\begin{widetext}
\begin{center}
\begin{table}
\caption{Summary of the results in the strongly interacting regime.}
\begin{tabular}{cccccc} 
\hline
~~$\frac{1}{a k_F}$~~ & ~~$\frac{\Delta}{E_{FG}}$~~ & ~~$\frac{E_{GFMC}}{A~E_{FG}}$~~
&~~$\frac{E_{mol}}{2~E_{FG}}$~~& ~~$\frac{R_{rms}}{r_0}$~~ 
& ~~$\mu R_{rms}$~~  \\
\hline
\hline
0 & 0.99(4) & 0.44(1)    &  0        &$\infty$ & $\infty$ \\
0.1 & 1.03(5) & 0.34(1)  & -0.01(1)  &  3.69 & 44.3\\
$0.\dot{3}$ & 1.37(5) & 0.02(1)  & -0.20(1) & 1.21  & 14.5\\
0.5         & 1.80(5) & -0.33(1) & -0.49(1) &  0.74 &  8.9 \\
1.0         & 3.2(1) & -2.23(1) & -2.31(1) & 0.38 &  4.6 \\
$1.\dot{3}$ & 5.7(3) & -4.58(2) & -4.63(1) &  0.28 &  3.4 \\
2.0         & 14.0(5) & -12.84(3) & -12.86(1) &  0.19 & 2.3 \\
\hline
\end{tabular}
\label{table_strong}
\end{table}
\end{center}
\end{widetext}


\clearpage

\begin{thebibliography}{}
\expandafter\ifx\csname natexlab\endcsname\relax\def\natexlab#1{#1}\fi
\expandafter\ifx\csname bibnamefont\endcsname\relax
  \def\bibnamefont#1{#1}\fi
\expandafter\ifx\csname bibfnamefont\endcsname\relax
  \def\bibfnamefont#1{#1}\fi
\expandafter\ifx\csname citenamefont\endcsname\relax
  \def\citenamefont#1{#1}\fi
\expandafter\ifx\csname url\endcsname\relax
  \def\url#1{\texttt{#1}}\fi
\expandafter\ifx\csname urlprefix\endcsname\relax\def\urlprefix{URL }\fi
\providecommand{\bibinfo}[2]{#2}
\providecommand{\eprint}[2][]{\url{#2}}

\bibitem[{\citenamefont{cooper}(1959)}]{cooper59}
\bibinfo{author}{\bibfnamefont{L.~N.} \bibnamefont{Cooper}},
 \bibinfo{author}{\bibfnamefont{R.~L.} \bibnamefont{Mills}},\bibnamefont{and}
  \bibinfo{author}{\bibfnamefont{A.~M.} \bibnamefont{Sessler}},
  \bibinfo{journal}{Phys. Review} \textbf{\bibinfo{volume}{114}},
  \bibinfo{pages}{1377} (\bibinfo{year}{1959}).

\bibitem[{\citenamefont{Pethick and Ravenhall}(1995)}]{pethick1995}
\bibinfo{author}{\bibfnamefont{C.~J.} \bibnamefont{Pethick}} \bibnamefont{and}
  \bibinfo{author}{\bibfnamefont{D.~G.} \bibnamefont{Ravenhall}},
  \bibinfo{journal}{Ann. Rev. Nuc. Part. Science}
  \textbf{\bibinfo{volume}{45}}, \bibinfo{pages}{429} (\bibinfo{year}{1995}).

\bibitem[{\citenamefont{Leggett}(1980)}]{leggett1980}
\bibinfo{author}{\bibfnamefont{A.~J.} \bibnamefont{Leggett}}, in
  \emph{\bibinfo{booktitle}{Modern Trends in the Theory of Condensed Matter}},
  edited by \bibinfo{editor}{\bibfnamefont{A.}~\bibnamefont{Pekalski}}
  \bibnamefont{and} \bibinfo{editor}{\bibfnamefont{R.}~\bibnamefont{Przystawa}}
  (\bibinfo{publisher}{Springer-Verlag}, \bibinfo{address}{Berlin},
  \bibinfo{year}{1980}).

\bibitem[{\citenamefont{Randeria}(1995)}]{randeria95}
\bibinfo{author}{\bibfnamefont{M.}~\bibnamefont{Randeria}}, in
  \emph{\bibinfo{booktitle}{Bose-Einstein Condensation}}, edited by
  \bibinfo{editor}{\bibfnamefont{A.}~\bibnamefont{Griffin}},
  \bibinfo{editor}{\bibfnamefont{D.}~\bibnamefont{Snoke}}, \bibnamefont{and}
  \bibinfo{editor}{\bibfnamefont{S.}~\bibnamefont{Stringari}}
  (\bibinfo{publisher}{Cambridge}, \bibinfo{year}{1995}).

\bibitem[{\citenamefont{Gorkov and Melik-Barkhudarov}(1961)}]{gorkov61}
\bibinfo{author}{\bibfnamefont{L.~P.} \bibnamefont{Gorkov}} \bibnamefont{and}
  \bibinfo{author}{\bibfnamefont{T.~K.} \bibnamefont{Melik-Barkhudarov}},
  \bibinfo{journal}{Sov. Phys. JETP} \textbf{\bibinfo{volume}{13}},
  \bibinfo{pages}{1018} (\bibinfo{year}{1961}).

\bibitem[{\citenamefont{De Marco et~al.}(1999)\citenamefont{De Marco}}]{demarco1999}
\bibinfo{author}{\bibfnamefont{B.} \bibnamefont{De~Marco}},
  \bibnamefont{and} 
  \bibinfo{author}{\bibfnamefont{D.~S.} \bibnamefont{Jin}},
 \bibinfo{journal}{Science}
  \textbf{\bibinfo{volume}{285}}, \bibinfo{pages}{1703} (\bibinfo{year}{1999}).

\bibitem[{\citenamefont{O'Hara et~al.}(2002)\citenamefont{O'Hara, Hemmer, Gehm,
  Granade, and Thomas}}]{ohara2002}
\bibinfo{author}{\bibfnamefont{K.~M.} \bibnamefont{O'Hara}},
  \bibinfo{author}{\bibfnamefont{S.~L.} \bibnamefont{Hemmer}},
  \bibinfo{author}{\bibfnamefont{M.~E.} \bibnamefont{Gehm}},
  \bibinfo{author}{\bibfnamefont{S.~R.} \bibnamefont{Granade}},
  \bibnamefont{and} \bibinfo{author}{\bibfnamefont{J.~E.}
  \bibnamefont{Thomas}}, \bibinfo{journal}{Science}
  \textbf{\bibinfo{volume}{298}}, \bibinfo{pages}{2179} (\bibinfo{year}{2002}).

\bibitem[{\citenamefont{Stenger et~al.}(1999)\citenamefont{Stenger, Inouye,
  Andrews, Miesner, Stamper-Kurn, and Ketterle}}]{stenger1999}
\bibinfo{author}{\bibfnamefont{J.}~\bibnamefont{Stenger}},
  \bibinfo{author}{\bibfnamefont{S.}~\bibnamefont{Inouye}},
  \bibinfo{author}{\bibfnamefont{M.~R.} \bibnamefont{Andrews}},
  \bibinfo{author}{\bibfnamefont{H.-J.} \bibnamefont{Miesner}},
  \bibinfo{author}{\bibfnamefont{D.~M.} \bibnamefont{Stamper-Kurn}},
  \bibnamefont{and} \bibinfo{author}{\bibfnamefont{W.}~\bibnamefont{Ketterle}},
  \bibinfo{journal}{Phys. Rev. Lett.} \textbf{\bibinfo{volume}{82}},
  \bibinfo{pages}{2422} (\bibinfo{year}{1999}).

\bibitem[{\citenamefont{Roberts et~al.}(2001)\citenamefont{Roberts, Claussen,
  Cornish, Donley, Cornell, and Wieman}}]{roberts2001}
\bibinfo{author}{\bibfnamefont{J.~L.} \bibnamefont{Roberts}},
  \bibinfo{author}{\bibfnamefont{N.~R.} \bibnamefont{Claussen}},
  \bibinfo{author}{\bibfnamefont{S.~L.} \bibnamefont{Cornish}},
  \bibinfo{author}{\bibfnamefont{E.~A.} \bibnamefont{Donley}},
  \bibinfo{author}{\bibfnamefont{E.~A.} \bibnamefont{Cornell}},
  \bibnamefont{and} \bibinfo{author}{\bibfnamefont{C.~E.}
  \bibnamefont{Wieman}}, \bibinfo{journal}{Phys. Rev. Lett.}
  \textbf{\bibinfo{volume}{86}}, \bibinfo{pages}{4211} (\bibinfo{year}{2001}).

\bibitem[{\citenamefont{Regal et~al.}(2003)\citenamefont{Regal, Ticknor,
  Bohn, and Jin}}]{regal2003}
\bibinfo{author}{\bibfnamefont{C.~A.} \bibnamefont{Regal}},
\bibinfo{author}{\bibfnamefont{C.} \bibnamefont{Ticknor}},
\bibinfo{author}{\bibfnamefont{J.~L.} \bibnamefont{Bohn}}, \bibnamefont{and} 
\bibinfo{author}{\bibfnamefont{D.~S.} \bibnamefont{Jin}},
  \bibinfo{journal}{Nature} \textbf{\bibinfo{volume}{424}},
  \bibinfo{pages}{47} (\bibinfo{year}{2003}).

\bibitem[{\citenamefont{Regal et~al.}(2004)\citenamefont{Regal, Greiner,
 and Jin}}]{regal2004}
\bibinfo{author}{\bibfnamefont{C.~A.} \bibnamefont{Regal}},
\bibinfo{author}{\bibfnamefont{M.} \bibnamefont{Greiner}}, \bibnamefont{and} 
\bibinfo{author}{\bibfnamefont{D.~S.} \bibnamefont{Jin}},
  \bibinfo{journal}{Phys. Rev. Lett.} \textbf{\bibinfo{volume}{92}},
  \bibinfo{pages}{040403} (\bibinfo{year}{2004}).

\bibitem[{\citenamefont{Lenz}(1929)}]{lenz1929}
\bibinfo{author}{\bibfnamefont{W.} \bibnamefont{Lenz}},
  \bibinfo{journal}{Z. Physik} \textbf{\bibinfo{volume}{56}},
  \bibinfo{pages}{778} (\bibinfo{year}{1929}).
  
\bibitem[{\citenamefont{Huang}(1957)}]{huang1957}
\bibinfo{author}{\bibfnamefont{K.} \bibnamefont{Huang}},  \bibnamefont{and} 
\bibinfo{author}{\bibfnamefont{C.~N.} \bibnamefont{Yang}},
  \bibinfo{journal}{Phys. Rev.} \textbf{\bibinfo{volume}{105}},
  \bibinfo{pages}{767} (\bibinfo{year}{1957}).  

\bibitem[{\citenamefont{Baker}(1999)}]{baker1999}
\bibinfo{author}{\bibfnamefont{G.~A.} \bibnamefont{Baker}},
  \bibinfo{journal}{Phys. Rev. C} \textbf{\bibinfo{volume}{60}},
  \bibinfo{pages}{054311} (\bibinfo{year}{1999}).

\bibitem[{\citenamefont{Heiselberg}(2001)}]{heiselberg2001}
\bibinfo{author}{\bibfnamefont{H.}~\bibnamefont{Heiselberg}},
  \bibinfo{journal}{Phys. Rev. A} \textbf{\bibinfo{volume}{63}},
  \bibinfo{pages}{043606} (\bibinfo{year}{2001}).

\bibitem[{\citenamefont{Carlson et~al.}(2003)\citenamefont{Carlson, Chang,
 Schmidt and Pandharipande}}]{carlson2003}
\bibinfo{author}{\bibfnamefont{J.} \bibnamefont{Carlson}},
  \bibinfo{author}{\bibfnamefont{S.-Y.}~\bibnamefont{Chang}},
  \bibinfo{author}{\bibfnamefont{V.~R.}~\bibnamefont{Pandharipande}}, 
   \bibnamefont{and}
   \bibinfo{author}{\bibfnamefont{K.~E.}~\bibnamefont{Schmidt}},
  \bibinfo{journal}{Phys. Rev. Lett.} \textbf{\bibinfo{volume}{91}},
  \bibinfo{pages}{50401} (\bibinfo{year}{2003}).


\bibitem[{\citenamefont{Pandharipande et~al.}(1973)}]{vijay73}
\bibinfo{author}{\bibfnamefont{V.~R.} \bibnamefont{Pandharipande}},
  \bibnamefont{and} \bibinfo{author}{\bibfnamefont{H.~A.}
  \bibnamefont{Bethe}},
  \bibinfo{journal}{Phys. Rev. C} \textbf{\bibinfo{volume}{7}},
  \bibinfo{pages}{1312} (\bibinfo{year}{1973}).

\bibitem[{\citenamefont{Cowell et~al.}(2002)\citenamefont{Cowell, Heiselberg,
  Mazets, Morales, Pandharipande, and Pethick}}]{cowell02}
\bibinfo{author}{\bibfnamefont{S.}~\bibnamefont{Cowell}},
  \bibinfo{author}{\bibfnamefont{H.}~\bibnamefont{Heiselberg}},
  \bibinfo{author}{\bibfnamefont{I.~E.} \bibnamefont{Mazets}},
  \bibinfo{author}{\bibfnamefont{J.}~\bibnamefont{Morales}},
  \bibinfo{author}{\bibfnamefont{V.~R.} \bibnamefont{Pandharipande}},
  \bibnamefont{and} \bibinfo{author}{\bibfnamefont{C.~J.}
  \bibnamefont{Pethick}}, \bibinfo{journal}{Phys. Rev. Lett.}
  \textbf{\bibinfo{volume}{88}}, \bibinfo{pages}{210403}
  (\bibinfo{year}{2002}).

\bibitem[{\citenamefont{Schmidt et~al.}(1977)}]{kevin77}
\bibinfo{author}{\bibfnamefont{V.~R.} \bibnamefont{Pandharipande}},
  \bibnamefont{and} \bibinfo{author}{\bibfnamefont{K.~E.}
  \bibnamefont{Schmidt}},
  \bibinfo{journal}{Phys. Rev. A} \textbf{\bibinfo{volume}{15}},
  \bibinfo{pages}{2486} (\bibinfo{year}{1977}).

   
\bibitem[{\citenamefont{Kalos}(1974)}]{kalos74}
\bibinfo{author}{\bibfnamefont{M.~H.} \bibnamefont{Kalos}},
\bibinfo{author}{\bibfnamefont{D.} \bibnamefont{Levesque}},
 \bibnamefont{and}
\bibinfo{author}{\bibfnamefont{L.} \bibnamefont{Verlet}},
  \bibinfo{journal}{Phys. Rev. A} \textbf{\bibinfo{volume}{9}},
  \bibinfo{pages}{2178} (\bibinfo{year}{1974}).

\bibitem[{\citenamefont{Anderson}(1975)}]{anderson75}
\bibinfo{author}{\bibfnamefont{J.~B.} \bibnamefont{Anderson}},
  \bibinfo{journal}{J. Chem. Phys.} \textbf{\bibinfo{volume}{63}},
  \bibinfo{pages}{1499} (\bibinfo{year}{1975}).

\bibitem[{\citenamefont{Bouchaud et~al.}(1988)\citenamefont{Bouchaud, Georges,
  and Lhuillier}}]{bouchaud1988}
\bibinfo{author}{\bibfnamefont{J.~P.} \bibnamefont{Bouchaud}},
  \bibinfo{author}{\bibfnamefont{A.}~\bibnamefont{Georges}}, \bibnamefont{and}
  \bibinfo{author}{\bibfnamefont{C.}~\bibnamefont{Lhuillier}},
  \bibinfo{journal}{J. Physique} \textbf{\bibinfo{volume}{49}},
  \bibinfo{pages}{553} (\bibinfo{year}{1988}).

\bibitem[{\citenamefont{petrov et~al.}(2003)\citenamefont{petrov}}]{petrov03}
\bibinfo{author}{\bibfnamefont{D.~S.}~\bibnamefont{Petrov}},
\bibfnamefont{C. \bibnamefont{Salomon}},
   \bibnamefont{and}
  \bibinfo{author}{\bibfnamefont{G.~V.}~\bibnamefont{Shlyapnikov}},
  \bibinfo{journal}{cond-mat/0309010} (\bibinfo{year}{2003}).
  
\bibitem[{\citenamefont{shumway et~al.}(2000)\citenamefont{shumway}}]{shumway00}
\bibinfo{author}{\bibfnamefont{J.} \bibnamefont{Shumway}},
   \bibnamefont{and}
  \bibinfo{author}{\bibfnamefont{D.~M.}~\bibnamefont{Ceperley}},
  \bibinfo{journal}{Journ. de Phys.}
  \textbf{\bibinfo{volume}{ IV 10 (P5)}},
  \bibinfo{pages}{3-16} (\bibinfo{year}{2000}).

\bibitem[{\citenamefont{Bartenstein et~al.}(2003)\citenamefont{Bartenstein, Altmeyer,
 Riedl, Jochim, Chin, Denschlag and Grimm}}]{bartenstein2004}
\bibinfo{author}{\bibfnamefont{M.} \bibnamefont{Bartenstein}},
  \bibinfo{author}{\bibfnamefont{A.}~\bibnamefont{Altmeyer}},
  \bibinfo{author}{\bibfnamefont{S.}~\bibnamefont{Riedl}}, 
  \bibinfo{author}{\bibfnamefont{S.}~\bibnamefont{Jochim}}, 
  \bibinfo{author}{\bibfnamefont{C.}~\bibnamefont{Chin}}, 
  \bibinfo{author}{\bibfnamefont{J.~H.}~\bibnamefont{Denschlag}}, 
   \bibnamefont{and}
   \bibinfo{author}{\bibfnamefont{R.}~\bibnamefont{Grimm}},
  \bibinfo{journal}{Phys. Rev. Lett.} \textbf{\bibinfo{volume}{92}},
  \bibinfo{pages}{120401-1} (\bibinfo{year}{2004}).

\end{thebibliography}
\end{document}